\documentclass{IEEEtran}
\IEEEoverridecommandlockouts
\usepackage{times,amsmath,amsfonts,amssymb,amsthm}
\usepackage{cite}
\usepackage{graphicx,subfigure,psfrag}
\usepackage{algorithm}%,algorithmic}
\usepackage{booktabs}
\usepackage{multirow}
\usepackage{multicol}
\usepackage{colortbl}
\usepackage{color}
\usepackage{algpseudocode}
\usepackage{balance}
\def\abf{{\bf a}}

\def\dbf{{\bf d}}

\def\pbf{{\bf p}}
\def\qbf{{\bf q}}
\def\rbf{{\bf r}}

\def\wbf{{\bf w}}
\def\xbf{{\bf x}}

\def\rbf{{\bf r}}
\def\xbf{{\bf x}}

\def\Fbf{{\bf F}}

\def\Ac{{\cal A}}

\def\Cc{{\cal C}}
\def\Dc{{\cal D}}

\def\Ic{{\cal I}}

\def\Kc{{\cal K}}

\def\Nc{{\cal N}}

\def\Qc{{\cal Q}}

\def\Sc{{\cal S}}
\def\Tc{{\cal T}}
\def\Uc{{\cal U}}

\def\ie{{\it i.e.,\ \/}}
\def\nn{\nonumber}
\def\tcb{\textcolor{blue}}

\theoremstyle{definition}

\newtheorem{theorem}{Theorem}

\newtheorem{proposition}{Proposition}
\newtheorem{myDef}{Definition}
\newtheorem{corollary}{Corollary}
\newtheorem{Remark}{Remark}

\begin{document}

\title{\huge Decentralized Caching under Nonuniform File Popularity and Size: Memory-Rate Tradeoff Characterization}

 \author{Yong~Deng, \IEEEmembership{Member, IEEE}  and Min~Dong, \IEEEmembership{Senior Member, IEEE}
\thanks{This work was supported by the Natural Sciences and Engineering Research Council of Canada (NSERC) under the Discovery Grant. Part of this work was presented in\cite{Deng2021Memory}. (Corresponding author: Min Dong.) }
\thanks{Yong Deng was with the Edward S. Rogers Sr. Department of Electrical and Computer Engineering,
University of Toronto, Toronto, ON M5S 3G4, Canada. He is now with the Department of Software Engineering, Lakehead University, Thunder Bay, ON P7B 5E1, Canada (email:
yong.deng@utoronto.ca).}
\thanks{Min Dong is with the Department of Electrical, Computer and Software Engineering, Ontario Tech University, Oshawa, ON L1G 0C5, Canada (e-mail: min.dong@ontariotechu.ca).}
}
\maketitle

\begin{abstract}
This paper aims to characterize the memory-rate tradeoff for decentralized caching  under nonuniform file popularity and size. We consider a  recently proposed decentralized modified coded caching scheme (D-MCCS) and formulate the cache placement optimization problem  to minimize the average rate for the D-MCCS. To solve  this challenging non-convex optimization problem, we first propose a successive Geometric Programming (GP) approximation algorithm, which guarantees convergence to a stationary point but has high computational complexity. Next, we develop a low-complexity file-group-based approach, where we propose a popularity-first and size-aware (PF-SA) cache placement strategy to partition files into two groups, taking into account the nonuniformity in file popularity and size. Both algorithms do not require the knowledge of active users  beforehand for cache placement. Numerical results show that they perform very closely to each other.  We further develop a lower bound  for decentralized caching under nonuniform file popularity and size  as a non-convex optimization problem and solved it using a similar successive GP approximation algorithm. We show that the  D-MCCS with  the optimized cache placement attains this lower bound when no more than two active users request files at a time. The same is true for files  with uniform size but  nonuniform  popularity and the optimal cache placement being  symmetric among files. In these cases, the optimized D-MCCS characterizes the exact memory-rate tradeoff for decentralized caching. For general cases, our numerical results show  that the average rate achieved by the optimized D-MCCS is very close to the lower bound.
\end{abstract}

%%%%%%%%%%%%%%%%%%%%%%%%%%%%%%%%%%%%%%%%%%%%
\section{Introduction}

Data caching at the network edge is anticipated to become a key technique to alleviate network congestion and reduce content delivery delay for future wireless networks~\cite{paschos2018role}.
 Coded caching that combines
an uncoded cache placement and a coded multicast delivery strategy has been proposed to harvest the global caching gain \cite{Maddah-Ali&Niesen:TIT2014}. The scheme has been shown to substantially reduce the delivery rate (load) as compared with uncoded caching. This promising result has attracted  significant interest in studying coded caching for different systems or network structures \cite{shariatpanahi2016multi,karamchandani2016hierarchical,pedarsani2016online,ji2016fundamental,Sengupta&etal:TIT17,Ibrahim2019Coded,Cao2019TCOM,Deng&Dong:Asilomar19}.

Many existing studies of coded caching generally rely on a centrally coordinated cache placement strategy  carefully designed to store a portion of each file at different subsets of users.
However, a coordinated cache placement may not always be possible, limiting the practical application of coded caching. Following this, decentralized caching has been considered \cite{Niesen&Maddah-Ali:TIT2015}, where   each user caches uncoded contents independently, requiring no coordination among users. For a system with a central server connecting to multiple cache-equipped users,  a \emph{decentralized coded caching scheme} (D-CCS) has been proposed in  \cite{Niesen&Maddah-Ali:TIT2015}, which consists of a decentralized (uncoded) cache placement  strategy and a coded delivery strategy.
Interestingly, for uniform  file popularity and size,  the performance of the D-CCS is shown to be close to that of the centralized coded caching scheme~\cite{Niesen&Maddah-Ali:TIT2015}. The D-CCS has since attracted many interests, with extensions to the system with nonuniform cache sizes~\cite{Mohammadi17Decentralized}, nonuniform file popularity~\cite{Niesen&Maddah-Ali:TIT2017,Ji&Order:TIT17,Zhang&Coded:TIT18} or sizes~\cite{Zhang&Lin15Coded,Zhang&Lin19Closing,Wang2019Optimization}.

For files with either nonuniform popularity or nonuniform  sizes,  cache placement  for the D-CCS have been studied  to  lower the average delivery rate. In particular, for files with nonuniform  popularity only,  a popularity-first (PF) strategy that allocates more cache to a more popular file has been considered in the cache placement design ~\cite{Niesen&Maddah-Ali:TIT2017,Ji&Order:TIT17,Zhang&Coded:TIT18},   while for files with nonuniform sizes only, a size-first (SF) strategy has been proposed that allocates more cache to a larger file~\cite{Zhang&Lin15Coded,Zhang&Lin19Closing}. These strategies are
designed based on one type of nonuniformity while ignoring
the other.  In~\cite{Wang2019Optimization}, a cache placement optimization problem for the D-CCS under nonuniform file popularity and
sizes was formulated and solved via numerical methods. 
Different lower bounds  for caching with any (coded or uncoded) placement have  been developed to quantify the performance of these proposed schemes. When  the server knows the active users (\ie users who request files) in advance for the cache placement, it has been shown that the achievable rate of the D-CCS over the tightest lower bound known in the literature is within a constant factor~\cite{Zhang&Coded:TIT18,Zhang&Lin19Closing}. However,    since these lower bounds are developed for centralized caching with any coded or uncoded cache placement, they are rather loose for decentralized caching. Also, the D-CCS is a suboptimal caching scheme. Thus, the gap between the D-CCS and the lower bounds is still large for practical consideration.
As a result, these existing results
\cite{Wang2019Optimization,Niesen&Maddah-Ali:TIT2017,Ji&Order:TIT17,Zhang&Coded:TIT18,Zhang&Lin15Coded,Zhang&Lin19Closing} are insufficient to characterize the
memory-rate tradeoff for decentralized caching under nonuniform file popularity or size. In particular, other than~\cite{Niesen&Maddah-Ali:TIT2017} and~\cite{Wang2019Optimization}, the above works cannot be applied to the scenario where the server does not know  the active users  in advance for cache placement.

Recently,  a \emph{decentralized modified coded caching scheme} (D-MCCS) has been proposed under uniform file popularity and size, assuming the server knows the active users in advance \cite{Yu&Maddah-Ali:TIT2018}. It improves upon the D-CCS by eliminating the redundant coded messages  in the D-CCS to further reduce the delivery rate. This scheme  has been shown to attain the lower bound developed for decentralized caching for both average and peak rates and thus characterizes the exact memory-rate tradeoff under uniform file popularity and size\cite{Yu&Maddah-Ali:TIT2018}.
The study of cache placement  for the D-MCCS under nonuniform file popularity and/or size is scarce. Only recently,  ~\cite{Zheng2020Decentralized} considered files with nonuniform sizes and proposed a suboptimal cache placement strategy  based on file grouping for the D-MCCS. Except for this, there is no other work on optimizing the cache placement  for the  D-MCCS or studying how optimal the D-MCCS is for decentralized caching.
In general, the memory-rate tradeoff for decentralized caching remains unknown under nonuniform file popularity or size.
\vspace{-1em}

\subsection{Contributions}
In this paper, we aim to characterize the memory-rate tradeoff  for decentralized caching  under nonuniform file popularity and size. Focusing on the D-MCCS, we formulate the cache placement optimization problem  to minimize the average rate. To solve this challenging non-convex optimization problem, we first propose a successive  Geometric Programming (GP) approximation algorithm, which guarantees  convergence to a stationary point of the optimization problem. Due to the high computational complexity involved in this algorithm, we further develop a low-complexity file-group-based approach for an approximate solution. In particular,  we propose a popularity-first and size-aware (PF-SA) cache placement strategy. It partitions the files into two  groups based on popularity for cache placement and determines the cached amount of each file in the popular group that  captures the nonuniformity in both file popularity and size. Unlike many existing decentralized caching schemes, both of our proposed approximation algorithm and PF-SA strategy for decentralized cache placement do not require the knowledge of the active user set in advance at the server. \tcb{}

To study the memory-rate tradeoff for decentralized caching, we further propose a lower bound on the average rate for decentralized caching under nonuniform file popularity and size. This lower bound is developed under the decentralized cache placement, which is different from the existing lower bounds developed for  caching under any (coded or uncoded) cache placement ~\cite{Niesen&Maddah-Ali:TIT2017,Ji&Order:TIT17,Zhang&Coded:TIT18,Wang2019Optimization,Zhang&Lin15Coded,Zhang&Lin19Closing}. We present the lower bound as a non-convex optimization problem and develop a similar  successive  GP approximation algorithm to obtain a stationary point of the optimization problem. For the case when  no more than two active users request files at a time, we show that the  D-MCCS with the optimized cache placement attains our proposed lower bound; This indicates that the scheme is optimal for decentralized caching, characterizing the exact memory-rate tradeoff. For files with uniform size but  nonuniform  popularity, we also identify a condition of symmetric cache placement  for the optimized D-MCCS to attain the proposed lower bound.  

Numerical results show that the average rate achieved by the proposed  PF-SA-cache-placement-based strategy is very close to, often even lower than, that of the successive GP approximation algorithm, but with significantly lower computational complexity.
Furthermore, the PF-SA cache placement strategy substantially outperforms existing PF or SF strategies when files contain nonuniformity in both popularity and sizes. Our numerical results also show that the performance gap between the lower bound and the optimized D-MCCS via either the successive  GP approximation algorithm or the PF-SA cache placement strategy  is very small in general. This not only
demonstrates the near-optimal performance of the PF-SA
cache placement strategy, but also indicates that the optimized D-MCCS is a  near-optimal  decentralized caching scheme under nonuniform file popularity and size.    
\vspace{-.8em}

%%%%%%%%%%%%%%%%%%%%%%%%%%%%%%%%%%%%%%%%%%%%%%%%%%%%%%%%%%%%%%
\subsection{Organization and Notations}
The rest of the paper is organized as follows. In Section~\ref{sec:relatedwork} we discuss related works. Section~\ref{sec:model} presents the system model. Section~\ref{sec:D-MCCS_optimization} describes the cache placement and content delivery procedures for the D-MCCS under nonuniform file popularity and sizes. In Section~\ref{sec:D_MCCS_solution}, we formulate the cache placement optimization problem for the D-MCCS and propose two different algorithms to obtain the  solution. In Section~\ref{sec:lb}, we propose a lower bound  for decentralized caching. We then  characterize the memory-rate tradeoff  by comparing the optimized D-MCCS with the lower bound in some special cases. Numerical results are presented in  Section~\ref{sec:simu}, followed by the conclusion in Section~\ref{sec:conclusion}.

\emph{Notations:} The cardinality of set $\Sc$ is denoted by $|\Sc|$, and   the index set for $\Sc$ is defined by $\Ic_{|\Sc|}=\{1,\ldots,|\Sc|\}$. The size of file $W$ is denoted by $|W|$.  The bitwise "XOR" operation between two subfiles is denoted by
 $\oplus$. Notation  $\Ac\backslash\Sc$ denotes  subtracting the elements in set  $\Sc$ from set $\Ac$. Notation $\abf\succcurlyeq{\bf 0}$ means vector $\abf$ is element-wise non-negative. 
%Also, we extend the definition of ${K \choose l}$ and define ${K \choose l}=0$,  for  $l<0$ or $l>K$.

%%%%%%%%%%%%%%%%%%%%%%%%%%%%%%%%%%%%%%%%%%%%%%%%%%%%%%%%%%%%%%%%%%%%%%%%
\section{Related Works}\label{sec:relatedwork}

\subsection{Centralized Caching}
Many recent works have studied the memory-rate tradeoff for caching in wireless networks.
For centralized caching with uniform file popularity and size, the memory-rate tradeoff was studied in~\cite{Maddah-Ali&Niesen:TIT2014,Wan:TIT2020,Yu&Maddah-Ali:TIT2018}, where different coded caching schemes and lower bounds for the delivery rate were proposed. In particular, for uniform file popularity and size, ~\cite{Yu&Maddah-Ali:TIT2018}  characterized the exact memory-rate tradeoff  under uncoded placement for both the peak and average rates. The  heterogeneity in the caching system, including nonuniform file popularity, file size, or cache size, was also investigated in~\cite{Deng&Dong:TIT22,Deng&DongMCCS:TIT22,Daniel&Yu:TIT19,Saberali&Lampe:TIT20,Jin&Cui:Arxiv2018,ji2016fundamental,Cao2019TCOM,Ibrahim2019Coded,Ramakrishnan2015An,Wan2017Novel,Asghari:TCOM19,Deng&Dong:COMML23} for centralized caching. For nonuniform file popularity, the cache placement optimization for common coded caching schemes was considered~\cite{Daniel&Yu:TIT19,Saberali&Lampe:TIT20,Jin&Cui:Arxiv2018,Deng&Dong:TIT22,Deng&DongMCCS:TIT22}, and the optimal cache placement structure has been fully characterized~\cite{Deng&Dong:TIT22,Deng&DongMCCS:TIT22}.
Several works also proposed improved coded delivery schemes for a given cache placement to reduce the delivery rate~\cite{Ramakrishnan2015An,Wan2017Novel,Asghari:TCOM19,Deng&Dong:COMML23}.

\subsection{Decentralized Caching}
Decentralized caching  was first considered in~\cite{Niesen&Maddah-Ali:TIT2015}, where the D-CCS was proposed under uniform file popularity and sizes, and its performance has been shown to be close to that of the centralized coded caching scheme. Subsequently, the D-MCCS was proposed in~\cite{Yu&Maddah-Ali:TIT2018} to  remove the redundancy in the coded messages of the D-CCS during the delivery phase. Under uniform file popularity and size, this scheme is  optimal  in terms of both average and peak delivery rates and thus  characterizes the exact memory-rate tradeoff for decentralized caching \cite{Yu&Maddah-Ali:TIT2018}.

When  different types of nonuniformity exist among files, the design becomes more complicated and more challenging to analyze. In the following, we mainly discuss the related works in this scenario. The differences between  these works and our work are summarized in Table~\ref{table:compare}.

\subsubsection{Cache Placement for the D-CCS} 
For  nonuniform file popularity or size, the existing works mainly focus on studying the cache placement problem for the D-CCS to reduce the average delivery rate. For  nonuniform file popularity, \cite{Niesen&Maddah-Ali:TIT2017,Ji&Order:TIT17,Zhang&Coded:TIT18}  proposed different file-group-based suboptimal cache placement strategies for the D-CCS, where files are divided into different groups, and  the same cache placement is applied to  the files within the same group. Furthermore, the PF strategy is  adopted to assign cache to different groups, where a group containing more popular files is allocated with more cache. For  the scenario of nonuniform file size only, the file-group-based approach has also been applied to the cache placement~design for the D-CCS~\cite{Zhang&Lin19Closing,Zhang&Lin15Coded}, where different file grouping methods have been proposed. For cache allocation to different file groups, the SF strategy was adopted to allocate more cache to the  group containing larger files. In general, both PF and SF strategies have been shown to perform well for their respective nonuniform  popularity only case and nonuniform  size only case. However, both strategies are
designed based on one type of nonuniformity while ignoring
the other, limiting their performance when files have  nonuniformity in  popularity and size.

While the file-group-based approach simplifies the cache placement design, it does not distinguish files of different popularity or sizes  within the same group, as well as the coding opportunities for files among  different groups, leading to suboptimal performance. The optimization approach was adopted to study   the cache placement for the D-CCS under nonuniform file size~\cite{Cheng&Li17Optimal} and under nonuniform file popularity and sizes~\cite{Wang2019Optimization}.  In both works, numerical methods were devised to find a solution to the optimization problem. All the above mentioned works, except for~\cite{Niesen&Maddah-Ali:TIT2017} and~\cite{Wang2019Optimization}, assume the active user set is known at the server in advance for cache placement.

\subsubsection{Cache Placement for the D-MCCS} 
The cache placement for the D-MCCS is  more difficult to design or analyze than that of the D-CCS, due to the much more complicated delivery strategy. Only recently, the cache placement problem for the D-MCCS has been studied for the nonuniform file size only scenario  in~\cite{Zheng2020Decentralized}. A heuristic file partitioning and grouping strategy has been proposed to simplify the problem, where  the SF strategy is adopted to allocate more cache to the file group with larger files.
However, no lower bound is provided to evaluate the performance of the proposed strategy. To the best of our knowledge,  the cache placement for the D-MCCS under nonuniform file popularity, or under the most general case of   nonuniform file popularity and sizes, has not been studied. How optimal the D-MCCS is in these scenarios is  still unknown.
 
\subsubsection{Lower Bounds} 
For  nonuniform file popularity or size, different lower bounds on the delivery rate for caching have  been developed
to evaluate the proposed cache placement strategies for the D-CCS in~\cite{Niesen&Maddah-Ali:TIT2017,Ji&Order:TIT17,Zhang&Coded:TIT18,Wang2019Optimization,Zhang&Lin15Coded,Zhang&Lin19Closing}.
With the active user set known at the server at the cache placement phase,
the D-CCS has been shown to be order-optimal, where the performance gap 
is within a  constant factor of the lower bound,  for either the nonuniform
file popularity only scenario ~\cite{Ji&Order:TIT17,Zhang&Coded:TIT18}, or
the nonuniform file size only scenario ~\cite{Zhang&Lin15Coded,Zhang&Lin19Closing}.
Nonetheless, this gap is still quite large, as the  lower bounds  developed
under any cache placement are loose, and  the delivery scheme of the D-CCS is suboptimal.   For \cite{Niesen&Maddah-Ali:TIT2017}
and~\cite{Wang2019Optimization} that  do not require the knowledge of the
active user set,  the performance gap of the proposed approaches  could be even larger.

In general, the lower bounds proposed by existing works are under any cache placement. They cannot
be used to  characterize the memory-rate tradeoff for decentralized caching
under the nonuniform file popularity or sizes, particularly when the  active users are unknown at the server.
In general,
the memory-rate tradeoff for decentralized caching under either nonuniform
file popularity or size remains an open problem to be characterized.

Besides the above works,  decentralized coded caching
has also been extended to other heterogeneous scenarios or system setups, including heterogeneous user profiles~\cite{Chang&Wang2019ISIT,Wang&Peleato19ISIT}, correlated files~\cite{Hassanzadeh2020TIT}, nonuniform cache sizes~\cite{Wang2019Optimization}, heterogeneous distortion\cite{Yang18TIT,Hassanzadeh20TWCOM}, multi-antenna transmission and shared caches\cite{Parrinello2020Fundamental}. 
\begin{table*}[t]
\renewcommand{\arraystretch}{1.05}
\centering
\caption{Summary of existing Decentralized Coded Caching schemes}
{
\begin{tabular}{|c|c|c|c|c|c|}
\hline
\multicolumn{1}{|c|}{References} & File popularity& File size&Delivery scheme  & Cache placement strategy&Lower bound  \\ \hline\hline
[14], [15], [16]               & Nonuniform&Uniform&D-CCS     & Multiple or two file
groups  w/ PF&Any placement      \\ \hline
[17], [18]             & Uniform &Nonuniform &D-CCS     & Multiple file groups  w/SF & Any placement        \\ \hline
[19]               &Nonuniform&Nonuniform &D-CCS & Optimization method &Any
placement       \\ \hline
[21]             & Uniform &Nonuniform &D-MCCS     & Multiple file groups  w/
SF&N/A       \\ \hline
Our work               &Nonuniform&Nonuniform &D-MCCS & Optimization method; Two file groups w/ PF-SA&Decentralized placement       \\ \hline
\end{tabular}
}
\label{table:compare}
\end{table*}

\allowdisplaybreaks
%%%%%%%%%%%%%%%%%%%%%%%%%%%%%%%%%%%%%%%%%%%%%%%%%%%%%%%%%%
\section{System Model }\label{sec:model}

\begin{figure}[t]
  \psfrag{K Users}{\bf \tiny $K$ Users}
   \psfrag{Cache size M}{\bf \tiny Cache size $M$}
   \centering
   \includegraphics[scale=0.23]{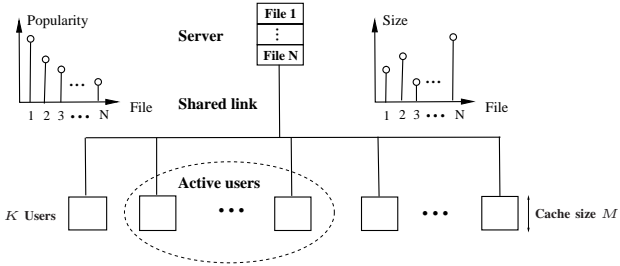}
   \caption{A cache-aided system with cache-equipped end users connecting to the server via a shared link.  The files in the server  have nonuniform  popularity and size. A set of unknown active users request files during the delivery phase. } \label{fig:sys_mod}\vspace{-1em}
 \end{figure}

We consider a cache-aided transmission system where a server connects to $K$ users over a shared error-free link, as shown in Fig~\ref{fig:sys_mod}. We denote the user index set by $\Kc\triangleq\{1,\ldots,K\}$.
Each user $k\in\Kc$  has a local cache of capacity $M$ bits.
The server has a database consisting of $N$ files, denoted as $W_1,\ldots,W_N$. We denote the file index set by $\Nc\triangleq\{1,\ldots,N\}$.
Each file $W_n,n\in\Nc$, is of size $F_n$ bits and has  probability  $p_n$ of being requested. We denote $\Fbf\triangleq[F_1,\ldots,F_N]^T$ as the file size vector, and denote $\pbf\triangleq[p_1,\ldots,p_N]^T$ as the popularity distribution of  all $N$ files, where $\sum_{n=1}^{N}p_n=1$.
{Without loss of generality, we sort
 the file indices as follows. First, we label the files according to the decreasing order of their popularity as $p_1\geq p_2\geq \cdots\geq p_N$. Next, for files with the same popularity but different sizes, we sort and label these files according to the decreasing order of their sizes; that is, if   $p_{n}=p_{n+1}$, we have $F_{n}\ge F_{n+1}$ for $n=1,.\ldots,N-1$. Files with the same popularity and size are ordered randomly.}   In this work, we only focus on the nontrivial case where the user's cache capacity is no greater than the total size of all $N$ files, \ie $M\le\sum_{n\in\Nc}F_n$.

The caching scheme operates in two phases: the cache placement phase and the content delivery phase.
In the cache placement phase, all users have access to the files stored in the server. For each file $n\in\Nc$, the users select a portion of its uncoded contents to store in their local caches.  With decentralized caching,
the cached contents are selected randomly by each user without any coordination among users. In the content delivery phase, a subset of users in $\Kc$ request files from the server. We refer to these users as \emph{active users}. Note that the server does not know these active users in advance in the cache placement phase. Let $p_{{\text a},k}$ denote the probability of user $k$ being active in the delivery phase. We define $\pbf_{\text a}\triangleq [p_{{\text a},1},\ldots,p_{{\text a},K}]^T$ as the probability  vector of users being active. Let $\Ac\subseteq\Kc$ denote the active user set. Let $d_k$ denote the index of the file requested by active user $k \in \Ac$. We define the demand vector of all the active users in $\Ac$ as $\dbf_{\Ac}\triangleq(d_k)_{k\in\Ac}$.
Based on the demand vector $\dbf_{\Ac}$ and the cached contents at users in $\Ac$, the server generates coded messages containing the uncached portion of requested files and transmits these messages to the active users.
Upon receiving the coded messages, each active user $k\in\Ac$  reconstructs its requested file $W_{\dbf_k}$ from the received coded messages and its own cached content. Note that, for a valid coded caching scheme,  each active user $k\in\Ac$ should be able to reconstruct its requested file $W_{d_{k}}$, for any demand vector $\dbf_\Ac$, assuming an error-free link.

\allowdisplaybreaks
%%%%%%%%%%%%%%%%%%%%%%%%%%%%%%%%%%%%%
\section{Decentralized Modified Coded Caching}\label{sec:D-MCCS_optimization}
In this section, we describe the cache placement and content delivery procedures of the D-MCCS for the system with nonuniform file popularity and size.

\subsection{Decentralized Cache Placement}\label{sec:MCCS_placement}
As mentioned earlier, the set of active users is unknown to the server before the content delivery phase. A salient feature of the decentralized caching considered in our work is that the cache placement strategy does not require the knowledge of the active user set $\Ac$ (both the size and the user identities) in the cache placement phase. We consider the following decentralized placement procedure: each user $k\in\Kc$ independently and randomly selects and caches $q_n F_n$ bits of file $W_n,n\in\Nc$, where $q_n$ is the fraction of $W_n$ the user wants to cache \ie
\begin{align}
0\leq q_n\leq 1,\quad n\in\Nc. \label{equ:de_cons2}
\end{align}

Following the common practice~\cite{Niesen&Maddah-Ali:TIT2015,Mohammadi17Decentralized,Wang2019Optimization,Niesen&Maddah-Ali:TIT2017,Ji&Order:TIT17,Zhang&Coded:TIT18,Zhang&Lin15Coded,Zhang&Lin19Closing}, we assume the file size $F_n$ is sufficiently large, such that $q_nF_n\in\mathbb{Z}$.\footnote{ The assumption of the file size $F_n$ being sufficiently large in terms of bits is reasonable in practice, since the file size typically exceeds $1$ kbit or even $1$ Mbits, which is usually large enough for $q_nF_n\in\mathbb{Z}$. } We define $\qbf\triangleq[q_1,\ldots,q_N]^T$ as the  cache placement vector for all the $N$ files. 
Since each file $W_n$ has $q_n F_n$ bits cached by each user of cache size $M$, we have the cache size constraint 
\begin{align}
\sum_{n=1}^{N}q_nF_n\leq M. \label{equ:de_cons1}
\end{align}
Note that the server knows the portion of each file cached  by each user $k\in\Kc$.

For uniform file popularity and size, \ie $p_1 = \cdots p_N= 1/N$ and $F_1=\cdots=F_N$, it has been shown that the symmetrical decentralized placement is optimal for the D-MCCS~\cite{Yu&Maddah-Ali:TIT2018}, \ie $q_1=\dots=q_N$.
For nonuniform file popularity and size, the cache placement may be different for different  files, which complicates the cache placement design for the D-MCCS. In this work, we aim to optimize the cache placement vector $\qbf$
for the D-MCCS to minimize the average delivery rate.
\vspace*{-.8em}

%%%%%%%%%%%%%%%%%%%%%%%%%%%%%%%%%%%%%%%%%%%%%%%%%%%%%%%%%%%%%%
\subsection{Content Delivery}\label{sec:delivery}
In the delivery phase, the server receives the information of the active user set $\Ac$ and their demand vector $\dbf_\Ac$. Based on these, the server knows the cached contents among the users in $\Ac$. We define subfile $W_{n,\Sc}$  as the chunk of  file $W_n$ that  is cached by the active user subset $\Sc\subseteq\Ac$ but not by the rest users in $\Ac$, \ie $\Ac\backslash\Sc$. We use $W_{n,\emptyset}$ to represent the portion of file $W_n$ that is not cached by any user in $\Ac$. Under the decentralized cache placement, if  file size $F_n, n\in\Nc$, is sufficiently large, by the law of large numbers, $q_n$ is approximately the probability of one bit in $W_n$ being selected and cached by a user. Following this, the size of subfile $W_{n,\Sc}$ is approximately given by~\cite{Niesen&Maddah-Ali:TIT2015}
\begin{align}
|W_{n,\Sc}|\approx q_{n}^{s}(1-q_{n})^{A-s}F_n,\quad\Sc\subseteq\Ac,|\Sc|=s\label{equ:de_subfile}
\end{align}
where $A \triangleq |\Ac|$. From \eqref{equ:de_subfile}, we note that besides $q_n$ and $F_n$, the size of subfile $W_{n,\Sc}$ also depends on $|\Sc|$, \ie the number of the users who cache this subfile.

 For any file demand vector $\dbf_\Ac$, the D-MCCS multicasts coded messages to  different user subsets in $\Ac$. Each coded message is intended for a unique active user subset $\Sc\subseteq\Ac$. It is formed by the bitwise XOR operation of total $|\Sc|$ subfiles, one from each requested file $d_k$ by user $k\in\Sc$, given by
\begin{align}
C_\Sc\triangleq\bigoplus_{k\in \Sc} \! W_{d_{k},\Sc\backslash\{k\}},\quad \Sc\subseteq\Ac,\Sc\neq \emptyset.\label{equ:de_codedmsg}
\end{align}
Note from \eqref{equ:de_codedmsg} that in $C_\Sc$, the subfile from each requested file $d_k$ by user $k$ is the one that is cached by users in $\Sc\backslash \{k\}$ exclusively.
Also note that the coded messages can only be formed for the nonempty active user subset $\Sc\neq\emptyset$.

For files with different popularities or sizes, the portion $q_nF_n$ of file $W_n$ cached by the users may be different for different files. As a result, the subfiles forming the coded message $\Cc_\Sc$ in \eqref{equ:de_codedmsg} may not have equal size. In this case, zero-padding is adopted for the XOR operation such that subfiles are zero-padded to the size of the largest subfile. Thus, the size of $C_\Sc$ is  determined by the largest subfile in $\Cc_\Sc$, \ie
\begin{align}
|C_\Sc|=&\max_{k\in\Sc}|W_{d_{k},\Sc\backslash \{k\}}|=\max_{k\in\Sc}q_{d_{k}}^{s}(1-q_{d_{k}})^{A-s}F_{d_k},\nn\\
&\quad\quad\quad\quad  \Sc\subseteq\Ac, |\Sc|=s+1, s=0,\ldots,A-1.\label{equ:de_msglength}
\end{align}

\begin{Remark}
For files with nonuniform file popularity or size, cache placement may be different for different files, resulting in subfiles of nonequal sizes, as shown in~\eqref{equ:de_subfile}. The existence of nonequal subfiles complicates the cache placement design. Zero-padding is a common technique to handle nonequal subfiles in formulating coded messages for coded caching in both centralized~\cite{Daniel&Yu:TIT19,Jin&Cui:Arxiv2018,Deng&Dong:TIT22} and decentralized~\cite{Wang2019Optimization} fashions. However, its impact on  decentralized coded caching has not been studied in the literature and is unknown. In Section~\ref{sec:tradeoff}, we will analyze in what scenarios using zero-padding incurs no loss of optimality. We will further use simulation to show the impact of zero-padding on the performance in Section~\ref{sec:simu}. 
\end{Remark}

In the original D-CCS~\cite{Niesen&Maddah-Ali:TIT2015}, for any file demand vector  $\dbf_{\Ac}$, the server transmits the coded messages corresponding to all the active user subsets $\{\Cc_\Sc:\forall\Sc\subseteq\Ac\}$ to the  active users. In contrast, for the D-MCCS, the server only transmits coded messages  corresponding to certain selected active user subsets~\cite{Yu&Maddah-Ali:TIT2018}. To describe the delivery procedure, we first provide the following two definitions:
\begin{myDef}\label{defLeader}
%\begin{enumerate}
\emph{Leader group}: For any demand vector $\dbf_{\Ac}$ containing $\widetilde{N}(\dbf_{\Ac})$ distinct requests,  the leader group $\Uc_{\Ac}$ is a subset of the active user set $\Ac$, with the following conditions hold: $\Uc_{\Ac}\subseteq \Ac$,  $|\Uc_{\Ac}|=\widetilde{N}(\dbf_{\Ac})$, and the users in $\Uc_{\Ac}$ have exactly $\widetilde{N}(\dbf_{\Ac})$ distinct requests.
\end{myDef}

\begin{myDef}\label{defRedundant}
\emph{Redundant group}: Given $\Uc_\Ac$, any active user subset $\Sc \subseteq \Ac$ is called a redundant group if $\Sc \cap \Uc_{\Ac}=\emptyset$; otherwise, $\Sc$ is a non-redundant group.
\end{myDef}
%\end{enumerate}

\begin{Remark}
Note that the leader group is not unique. When multiple users request  the same file, we only select one of these users to be in the leader group, and   there are multiple choices  to form it. The key feature of the leader group is that the files requested by users in
the leader group should be distinct and represent all files requested by
the active users. Also, once a leader group is formed, it should be kept  to carry out the coded delivery procedure.
\end{Remark}

The delivery procedure of the D-MCCS improves upon that of the D-CCS by multicasting only  coded messages corresponding to the non-redundant groups, \ie $\{C_\Sc: \forall \Sc\subseteq\Ac$ and $\Sc \cap \Uc_{\Ac}\neq\emptyset\}$, to both non-redundant and redundant groups.\footnote{
Note that this coded delivery strategy follows that of the centralized MCCS~\cite{Yu&Maddah-Ali:TIT2018}, which has been shown to be a valid strategy, \ie a user can reconstruct any requested file by using the strategy.
} As a result, the D-MCCS achieves a lower delivery rate than the D-CCS.  Note that the rate reduction only occurs when redundant groups exist, \ie there are multiple requests of the same file among the active users.

We summarize both the cache placement and the coded delivery procedures of the D-MCCS in Algorithm~\ref{Alg:D-MCCS}.
With the cached contents at each user via the decentralized cache placement described in Section~\ref{sec:MCCS_placement} and the coded messages $\{C_\Sc:\! \forall \Sc\subseteq\Ac~\text{and}~\Sc \cap \Uc_{\Ac}\neq\emptyset\}$ multicasted by the server, each user in $\Ac$ can retrieve all the subfiles required and reconstruct  its requested file~\cite{Yu&Maddah-Ali:TIT2018}.

\begin{algorithm}[t]
\caption{ Decentralized modified coded caching scheme}\label{Alg:D-MCCS}
\begin{algorithmic}[1]
\State Decentralized cache placement procedure:
\For{$n\in\Nc$}
    \State Each user $k\in\Kc$ randomly caches  $q_nF_n$ bits of file $W_n$.
\EndFor

\State Coded delivery procedure:
\For{$\Sc\subseteq\Ac$ and $\Sc \cap \Uc_{\Ac}\neq\emptyset$}
    \State The server generates $\Cc_\Sc$ based on~\eqref{equ:de_codedmsg} and multicasts it to $\Sc$.
\EndFor
\end{algorithmic}
\end{algorithm}

%%%%%%%%%%%%%%%%%%%%%%%%%%%%%%%%%%%%%%%%%%%%%%%%%%%%%%%%%%%%%%%%%%%%%
\section{Decentralized Cache Placement Optimization}\label{sec:D_MCCS_solution}
In this section, we first formulate the cache placement design for the D-MCCS under nonuniform file popularity and size into a cache placement optimization problem to minimize the average delivery rate. We then develop two algorithms to solve this optimization problem.
%\vspace*{-.8em}

%%%%%%%%%%%%%%%%%%%%%%%%%%%%%%%%%%%%%%%%%%%%%%%%%%
\subsection{Problem Formulation}
Based on the delivery procedure in the D-MCCS described in Section~\ref{sec:delivery}, for a given demand vector $\dbf_{\Ac}$, the delivery rate is the total number of bits in the coded messages corresponding to all the non-redundant groups $\{C_\Sc: \forall \Sc\subseteq\Ac$ and $\Sc \cap \Uc_{\Ac}\neq\emptyset\}$, expressed as
\begin{align}\label{DeliveryRate_0}
R_{\text{MCCS}}(\dbf_\Ac;\qbf)&=\sum_{\Sc\subseteq\Ac,\Sc\cap\Uc_\Ac\ne\emptyset} |C_\Sc|.
\end{align}
 Define $\Qc^{s}\triangleq \{\Sc\subseteq\Ac:\Sc\cap\Uc_\Ac\ne\emptyset, |\Sc|=s\}$ as the set of the non-redundant groups with $|\Sc|=s$ users, for $ s=1,\ldots,K$.  Based on the expression of $|\Cc_\Sc|$ in \eqref{equ:de_msglength}, we can rewrite \eqref{DeliveryRate_0} as
\begin{align}\label{CodedMsgPad}
R_{\text{MCCS}}(\dbf_\Ac;\qbf)=\sum_{s=0}^{A-1}\sum_{\Sc\in\Qc^{s+1}}\max_{k\in\Sc}q_{d_{k}}^{s}(1-q_{d_{k}})^{A-s}F_{d_k}.
\end{align}
By taking the expectation of $R_{\text{MCCS}}(\dbf_\Ac;\qbf)$ over all the possible $\dbf_\Ac\in\Nc^A$ and $\Ac\subseteq\Kc$, we obtain the average rate of the D-MCCS as
a function of $\qbf$ as
\begin{align}\label{equ:de_obj1}
\bar R_\text{MCCS}(\qbf)&=E_{\Ac}\left[E_{\dbf_\Ac}[R_{\text{MCCS}}(\dbf_\Ac;\qbf)]\right]\nn\\
&=E_{\Ac}\left[ \sum_{\dbf_{\Ac}\in\Nc^{A}}\left(\prod_{k\in\Ac}p_{d_{k}}\right)R_{\text{MCCS}}(\dbf_\Ac;\qbf)\right].
\end{align}
Thus, we formulate the cache placement optimization problem for the D-MCCS under nonuniform file popularity and size as
\begin{align}
\textrm{{\bf P0}}: \quad &\min_{\qbf} \bar R_\text{MCCS}(\qbf)\nn\\
&\ \text{s.t.}\quad \eqref{equ:de_cons2}, \eqref{equ:de_cons1}.\nn
\end{align}

Note that {\bf P0} is a non-convex optimization problem with respect to (w.r.t.) $\qbf$, which is difficult to solve. In the following subsections, we propose two algorithms to solve {\bf P0}.
We first develop an iterative algorithm to solve {\bf P0}, which is guaranteed to converge to a stationary point of {\bf P0}. To reduce the computational complexity, we further  propose a low-complexity heuristic approach to compute an approximate solution for {\bf P0}.

%%%%%%%%%%%%%%%%%%%%%%%%%%%%%%%%%%%
\subsection{Successive GP Approximation Algorithm}\label{sec:SP_MCCS}
To solve {\bf P0}, we first reformulate {\bf P0} into an equivalent Complementary GP (CGP) problem~\cite{avriel2013advances}. Then, we adopt the successive GP approximation method proposed in~\cite{Chiang2007Power} to find a solution for {\bf P0}.

To reformulate {\bf P0} into an equivalent CGP problem, we first introduce auxiliary variables $x_n, n\in\Nc$, and add the following inequality constraint for the term $(1-q_{n})$ in \eqref{CodedMsgPad}:
\begin{align}
1-q_n\leq x_n,\quad n\in\Nc.\label{equ:de_cons3}
\end{align}
We further introduce auxiliary variables $w_{\dbf_\Ac,\Sc}$, for $\dbf_\Ac\in\Nc^A$,  $\Ac\subseteq\Kc$, and $\Sc\in\Qc^{s+1}, s=0,\ldots,A-1$. In the expression of $R_{\text{MCCS}}(\dbf_\Ac;\qbf)$ in~\eqref{CodedMsgPad}, we replace $\max_{k\in\Sc}q_{d_k}^{s}(1-q_{d_k})^{A-s}F_{d_k}$ by $w_{\dbf_\Ac,\Sc}$, and based on \eqref{equ:de_cons3}, we add the following constraints
\begin{align}
q_{d_{k}}^{s}x_{d_{k}}^{A-s}&F_{d_k} \leq w_{\dbf_\Ac,\Sc},\quad k\in\Sc\label{equ:de_cons4}
\end{align}
for given $\Sc \in \Qc^{s+1}, \dbf_\Ac \subseteq \Nc^A, \Ac \subseteq \Kc$. With these auxiliary variables and constraints \eqref{equ:de_cons3} and \eqref{equ:de_cons4}, we can reformulate {\bf P0} into
the following equivalent problem:
\begin{align}
\textrm{\bf P1}:\! &\;\min_{\qbf\succcurlyeq0,\xbf\succcurlyeq0,\wbf\succcurlyeq0} \!\!E_{\Ac}\bigg[\sum_{\dbf_\Ac\in\Nc^A}\!\!\bigg(\prod_{k\in\Ac}p_{d_k}\!\!\bigg)\sum_{s=0}^{A-1}\sum_{\Sc\in\Qc^{s+1}}\!\!\!w_{\dbf_\Ac,\Sc}\bigg] \nn\\
\textrm{s.t.} &\;\; q_n\leq 1,\quad\quad\quad n\in\Nc,\label{equ:MCCS_cons_1}\\
 &\sum_{n=1}^{N}q_nF_nM^{-1}\leq 1,\label{equ:MCCS_cons_2}\\
 &\frac{1}{q_n+x_n}\leq 1,\quad\quad n\in\Nc,\label{equ:MCCS_cons_3}\\
 &w_{\dbf_\Ac,\Sc}^{-1}\cdot q_{d_{k}}^{s}x_{d_{k}}^{A-s}F_{d_k} \leq 1,\ k\in\Sc, \Sc\in\Qc^{s+1},  \nn\\
 &\quad\quad\quad\quad\quad\     s=0,\ldots,A-1,\dbf_\Ac\in\Nc^A,\Ac\subseteq\Kc \label{equ:MCCS_cons_4}
\end{align}
where $\xbf\triangleq[x_1,\ldots,x_N]^T$, and $\wbf\triangleq(w_{\dbf_\Ac ,\Sc})$ is the vector containing all  $w_{\dbf_\Ac,\Sc}$'s, for $\Sc\in\Qc^{s+1}, s=0,\ldots,A-1$ and $\dbf_\Ac\in\Nc^A$,  $\Ac\subseteq\Kc$.
Note that the optimization variables $\qbf,\xbf,\wbf$ are all nonnegative. Also, constraints \eqref{equ:MCCS_cons_2}, \eqref{equ:MCCS_cons_3} and \eqref{equ:MCCS_cons_4} are the re-expressions of constraints \eqref{equ:de_cons1},  \eqref{equ:de_cons3} and \eqref{equ:de_cons4}, respectively. 

In {\bf P1}, the objective function is a posynomial, and the constraint functions at the left hand side (LHS) of \eqref{equ:MCCS_cons_1},  \eqref{equ:MCCS_cons_2}, and \eqref{equ:MCCS_cons_4}  are also posynomials. Also, the constraint function at LHS of \eqref{equ:MCCS_cons_3} can be viewed as the ratio of two posynomials (\ie $1$ and $q_n+x_n$).
Thus, {\bf P1} is a CGP problem. A CGP problem is in general an intractable NP-hard problem~\cite{avriel2013advances}. A successive approximation  approach has been developed in~\cite{Chiang2007Power}, which uses a sequence of GP approximations to obtain a stationary point of the problem. We adopt this approach to solve {\bf P1}, where we compute $(\qbf, \xbf, \wbf)$ iteratively via a sequence of GP approximations.

Denote the objective function in {\bf P1} by $\bar{R}_\text{MCCS}^{\text{\tiny CGP}}(\!\qbf,\xbf,\wbf\!)$.
Let $(\qbf^{(i)},\xbf^{(i)},\wbf^{(i)})$ denote the solution obtained in iteration $i$.
In iteration $i+1$, given $(\qbf^{(i)},\xbf^{(i)})$, we form the following  GP approximation of {\bf P1}:
\begin{align}
\textrm{\bf P2}&\big(\qbf^{(i)}\!,\!\xbf^{(i)}\big):\min_{\qbf\succcurlyeq0,\xbf\succcurlyeq0,\wbf\succcurlyeq0}\!\bar R^{\text{\tiny CGP}}_{\text{MCCS}}(\!\qbf,\xbf,\wbf\!) \nn\\
& \textrm{s.t.}\;\; \eqref{equ:MCCS_cons_1},\eqref{equ:MCCS_cons_2},\eqref{equ:MCCS_cons_4},\nn\\
&\quad\frac{1}{\left(q_n^{(i)}\!+\!x_n^{(i)}\right) \left(\frac{q_n}{q_n^{(i)}}\!\! \right)^{\!\alpha_n^{(i)}}\! \left(\frac{x_n}{x_n^{(i)}} \!\right)^{\beta_n^{(i)}}}\leq1, \quad n\in\Nc \label{equ:MCCS_cons_5}
\end{align}
where $\alpha_n^{(i)}\triangleq\frac{q_n^{(i)}}{q_n^{(i)}+x_n^{(i)}}$ and $\beta_n^{(i)}\triangleq\frac{x_n^{(i)}}{q_n^{(i)}+x_n^{(i)}}$. Since the constraint function at LHS of~\eqref{equ:MCCS_cons_5} is a posynomial, $\textrm{\bf P2}\big(\qbf^{(i)}\!,\!\xbf^{(i)}\big)$ is a standard GP problem. Comparing constraints~\eqref{equ:MCCS_cons_3} and~\eqref{equ:MCCS_cons_5}, we note that based on the arithmetic-geometric mean inequality, we have~\cite{Chiang2007Power}
\begin{align}\label{equ:arithmetic}
q_n+x_n&\geq \bigg(\frac{q_n}{\alpha_n^{(i)}} \bigg)^{\alpha_n^{(i)}} \bigg(\frac{x_n}{\beta_n^{(i)}} \bigg)^{\beta_n^{(i)}}\nn\\
&=\left(q_n^{(i)}\!+\!x_n^{(i)}\right) \bigg(\frac{q_n}{q_n^{(i)}}\!\! \bigg)^{\!\alpha_n^{(i)}}\! \bigg(\frac{x_n}{x_n^{(i)}} \!\bigg)^{\beta_n^{(i)}}.
\end{align}
It follows that constraint \eqref{equ:MCCS_cons_5} in {\bf P2} tightens constraint \eqref{equ:MCCS_cons_3} in {\bf P1}. This guarantees that any solution to {\bf P2} is also a feasible solution to {\bf P1}.

Problem {\bf P2}$(\qbf^{(i)},\xbf^{(i)})$ can be solved using a standard convex solver. we obtain $(\qbf^{(i+1)},\xbf^{(i+1)}, \wbf^{(i+1)})$ as the optimal solution of {\bf P2}$(\qbf^{(i)},\xbf^{(i)})$. As shown in~\cite[Proposition 3]{Chiang2007Power}, the above approach of iteratively solving {\bf P2}$(\qbf^{(i)},\xbf^{(i)})$ is guaranteed to converge to a stationary point of {\bf P1}.
We summarize this successive GP approximation algorithm for {\bf P1} in Algorithm \ref{alg:MCCS1}.
By the equivalence of {\bf P0} and {\bf P1}, we can use Algorithm \ref{alg:MCCS1} to compute a stationary point of {\bf P0}.  

\emph{Complexity Analysis:} Note that {\bf P2}$(\qbf^{(i)},\xbf^{(i)})$ has
$\sum_{A=1}^{K}\binom{K}{A}N^A2^A+2N$ optimization variables and $\sum_{A=1}^{K}\binom{K}{A}N^A2^AA+2N+1$
constraints, which grow exponentially with $K$, the same for computing the objective function. A GP problem is typically solved by the interior
point method, whose complexity is in the polynomial time of the problem size. Thus, the overall complexity
grows exponentially with $K$. As a result, the computational complexity of Algorithm~\ref{alg:MCCS1} can be very high as the number of users $K$ increases. To address this issue, in the next subsection, we develop an alternative algorithm to provide an approximate solution to the problem with very low complexity.

\begin{algorithm}[t]
\caption{The successive GP approximation algorithm for {\bf P0}}\label{alg:MCCS1}
\begin{algorithmic}[1]
\Require
   { $K$,  $M$,  $N$, $\pbf$, $\pbf_{\text a}$.}
\Ensure
   {$\qbf^*$, $\bar R_{\text{MCCS}}^*$.}\\
{\bf Initialization:} Set initial feasible point $\left(\qbf^{(0)},\xbf^{(0)}, \wbf^{(0)}\right)$. Set $i=0$.
%\Ensure
\Repeat
    \State Solve {\bf P2}$(\qbf^{(i)},\xbf^{(i)})$ to obtain $\left(\qbf^{(i+1)},\xbf^{(i+1)}, \wbf^{(i+1)}\right)$.
    \State Set $i=i+1$.
\Until $\bar R^{\text{\tiny CGP}}_{\text{MCCS}}(\qbf^{(i)},\xbf^{(i)},\wbf^{(i)})$ converges.
\\
Set $\qbf^*=\qbf^{(i)}$; $\bar R_{\text{MCCS}}^{*}=\bar R^{\text{\tiny CGP}}_{\text{MCCS}}(\qbf^{(i)},\xbf^{(i)},\wbf^{(i)})$.
\end{algorithmic}
\end{algorithm}

%%%%%%%%%%%%%%%%%%%%%%%%%%%%%%%%%%%%%%%%%%%%%%%%%%%%%%%%%
\subsection{Low-Complexity File-Group-Based Approach}\label{sec:MCCS_low_complex}
To address the high computational complexity faced in GP approximation algorithm in Algorithm \ref{alg:MCCS1},  we propose a popularity-first and size-aware (PF-SA) cache placement strategy. It uses the file-grouping concept~\cite{Ji&Order:TIT17,Zhang&Coded:TIT18} to categorize files based on popularity into two groups for cache placement, and the cached amount of each file is size-dependent.  Using this strategy, we provide an  approximate solution for {\bf P0} with low computational complexity. Although file grouping has also been used for the cache placement strategies proposed in~\cite{Ji&Order:TIT17,Zhang&Coded:TIT18}, there are some differences of our strategy from those, which will be discussed in Remark~\ref{remark3}. 

\subsubsection{PF-SA Cache Placement}\label{sec:PF-SA} 
In the cache placement phase, we partition the $N$ files into two groups according to their popularity in $\pbf$. Define $\Nc_1\triangleq\{1,\ldots,N_1\}$, for $N_1\in\Nc$ and $\Nc_2\triangleq \Nc\backslash\Nc_1$ as the file index sets of the first and second file groups, respectively. Recall that the indices of files are ordered according to the decreasing order of their popularity. As a result, the first file group $\Nc_1$ contains the $N_1$ most popular files, and the second file group $\Nc_2$ contains the remaining unpopular files in $\Nc$. We allocate each user's entire cache of $M$ bits to the files in the first group $\Nc_1$. For these files in $\Nc_1$, regardless of their popularity and sizes, users randomly select and cache a portion of each file using the same fraction, \ie $q_1=\ldots=q_{N_1}$. The unpopular files in the second group $\Nc_2$ are not cached by any users and are solely stored at the server. Thus, for this two-file-group-based placement, the fraction of each file $n\in\Nc$ cached by any user $k\in\Kc$  is  given by
\begin{align}\label{equ:placement}
q_n=
\begin{cases}
\frac{M}{\sum_{n'\in\Nc_1}F_{n'}}, &\quad n\in \Nc_1,\\
0,&\quad n\in \Nc_2.
\end{cases}
\end{align}
Note that to ensure the entire cache memory is fully used, we always choose $N_1\in\Nc$ such that  $\sum_{n=1}^{N_1}F_{n}\ge M$.

As indicated above, in the proposed two-file-group-based placement, the $N_1$ most popular files are prioritized for the cache placement, and the entire user's cache is allocated to them. {Note that there may be multiple files with the same popularity.  Recall from Section II that in this case, we label file indices according to the decreasing order of their sizes. Thus, for the same popularity, a file of larger size is prioritized into the group of $N_1$ most popular files for cache placement.}  Also, note that although the same fraction of files is used for these $N_1$ files, the actual number of bits from file $W_n$ cached at each user  is $q_nF_n$, which depends on the file size $F_n$. Following this, a larger file will have more bits being stored at each user. Therefore, under this proposed placement, within the most popular file group, files of larger sizes are prioritized for cache placement. As a result, the salient feature of our proposed PF-SA cache placement strategy is that it captures the file nonuniformity in both popularity
and file size.
In contrast,  the existing popular heuristic cache placement strategies only prioritize files based on one  type of nonuniformity but ignore the other~\cite{Daniel&Yu:TIT19,Deng&DongMCCS:TIT22}, limiting their performance in the presence of nonuniformity in both popularity and size. We discuss them in the following remarks.

\begin{Remark}\label{remark:SFvsPF}
For both centralized and decentralized coded caching,  two types of strategies are considered for cache placement, \ie the PF strategy that allocates more cache to a more popular file or file group~under nonuniform file popularity \cite{Daniel&Yu:TIT19,Deng&DongMCCS:TIT22,Niesen&Maddah-Ali:TIT2017,Ji&Order:TIT17,Zhang&Coded:TIT18}, or the SF strategy that allocates more cache to a larger file or file group~under nonuniform file size\cite{Daniel&Yu:TIT19,Zhang&Lin15Coded,Zhang&Lin19Closing,Zheng2020Decentralized}.  However, both PF and SF strategies have their own drawbacks, as they are designed based on one type of nonuniformity while ignoring the other.
In particular, when files are  nonuniform in both popularity and sizes, the PF  strategy~\cite{Daniel&Yu:TIT19} specifies that  the number of bits from a less popular file cached by a user should be no more than that of a more popular file. The SF strategy is similarly defined, except that a more popular file is replaced by a larger file. Under such a restriction, for the PF strategy, there can be a scenario where a less popular large file may never have a chance to be cached by any user, even when the cache size is large enough to accommodate the files. As a result, the cache memory may not be fully used in some cases.  The existing numerical studies for the centralized caching scenarios have shown that the SF strategy tends to achieve a lower average rate  than that of the PF strategy~\cite{Daniel&Yu:TIT19,Deng&DongMCCS:TIT22}. However, the SF strategy ignores the differences in file popularity, which is an important indicator for caching. Thus, it may still be suboptimal, especially when the popular files are of relatively small size. 
 \end{Remark}
 
\begin{Remark}
Our proposed PF-SA cache placement takes into account the file nonuniformity in both  popularity and size to exploit the benefits of both PF and SF strategies for the cache placement design. This feature enables us to further exploit the caching gain in the scenario of nonuniform file popularity and sizes. Note that between file popularity and size, the PF-SA cache placement strategy puts a higher priority on file popularity as it is used to determine whether a file will be cached or not. Only among popular files, a larger file will be given more cache allocation. In the simulation, we will compare our PF-SA cache placement strategy with both the PF and SF strategies using two file groups. Note that for the SF strategy with two file groups, files are indexed in the decreasing order of their sizes; then, the $N_1$ largest files are placed into the first file group $\Nc_1$, and the rest files are in the second file group $\Nc_2$.
The simulation results show that the PF-SA cache placement always leads to the lowest average rate. Furthermore, its achieved average rate is very close to the lower bound developed in Section~\ref{sec:lb}, indicating that the PF-SA cache placement is near-optimal for the D-MCCS.  
\end{Remark}

\subsubsection{Optimization under PF-SA Cache Placement}
Under the PF-SA cache placement in~\eqref{equ:placement}, we can rewrite the average rate of the D-MCCS in~\eqref{equ:de_obj1} as a function of $N_1$. Following this, we reformulate {\bf P0} into  an optimization problem w.r.t. $N_1$ to minimize the average rate. Let $\Ac_1$ and $\Ac_2$ denote the sets of active users  who request the files in $\Nc_1$ and $\Nc_2$, respectively. Note that $\Ac_1 \cap \Ac_2 = \emptyset$ and $\Ac=\Ac_1\cup\Ac_2$. Denote $A_i=|\Ac_i|$, for $i=1,2$. Also denote $\dbf_{\Ac_i}$ as the files requested by users in $\Ac_i$ for $i=1,2$. Accordingly, the number of distinct file requests from $\Ac_i$ is $\widetilde N(\dbf_{\Ac_i})$.
Note that $\Nc_i$, $\Ac_i$, $\dbf_{\Ac_i}$ and $\widetilde N(\dbf_{\Ac_i})$, $i=1,2$ are all functions of $N_1$.  The reformulated problem is stated in the following proposition.

\begin{proposition}\label{pro:upper}
Consider the decentralized caching problem of $N$ files with popularity distribution $\pbf$ and sizes $\{F_i\}$, and $K$ users each with cache size $M$ bits and with probability $p_{\text{a},k}$ being active. The file indices are labelled according to the decreasing order of the file popularity. The minimum average rate of the D-MCCS under the PF-SA cache placement is
\begin{align}
\min_{N_1\in\Nc} \bar R^{\text{\tiny PF-SA}}_{\text{MCCS}}(N_1)\label{equ_min_two_group}
\end{align}where
$\bar R^{\text{\tiny PF-SA}}_{\text{MCCS}}(N_1)$ is  given by
\begin{align}\label{equ:prop_1}
\!\!\bar R^{\text{\tiny PF-SA}}_{\text{MCCS}}(N_1)\triangleq E_{\Ac}\!\bigg[\! \sum_{\dbf_{\Ac}\in\Nc^{A}}\!\!\bigg(\prod_{k\in\Ac}p_{d_{k}}\bigg)R^{\text{\tiny PF-SA}}_{\text{MCCS}}(\dbf_\Ac;N_1) \bigg]
\end{align} 
with $R^{\text{\tiny PF-SA}}_{\text{MCCS}}(\dbf_\Ac;N_1)$ being the delivery rate under the PF-SA cache placement for given $\dbf_\Ac$ and $N_1$, expressed as
\begin{align}\label{equ:two_groups_rate}
&R^{\text{\tiny PF-SA}}_{\text{MCCS}}(\dbf_\Ac;N_1)= \sum_{s=0}^{A-1}\sum_{\substack{\Sc\in\Qc^{s+1}\\ \Sc\cap\Ac_1\ne \emptyset}}\!\!\bigg(\! \frac{M}{\sum_{n\in\Nc_1}\!F_n}\!\bigg)^{s}\nn\\
&\quad\quad\quad\cdot \bigg(1- \frac{M}{\sum_{n\in\Nc_1}F_n}\bigg)^{A-s}\max_{k\in\Sc}F_{d_k} + \!\!\sum_{n\in\dbf_{\Ac_2}}\!\!F_n
\end{align} 
where $A=|\Ac|$.
\end{proposition}
\IEEEproof
See Appendix~\ref{proof_pro:upper}.
\endIEEEproof

\begin{algorithm}[t]
\caption{The PF-SA-cache-placement-based approximate solution for {\bf P0}}\label{Alg:Approx_solu}
\begin{algorithmic}[1]
\Require
   { $K$,  $M$,  $N$, $\pbf$, $\pbf_{\text{a}}$.}
%\State Sort the files in a specific order (either PF or SF) 
\Ensure
    {$\qbf^*$, $\bar R^{\text{\tiny FG-2}}_{\text{MCCS}}(N_1^*)$.}  
\For{$N_1$ = $1$ to $N$}
    %\State Set $q_1=\ldots=q_{N_1}=M/N_1$;
    %\State Set $q_{N_1+1}=\ldots=q_{N}=0$.
\State Compute $\bar R^{\text{\tiny PF-SA}}_{\text{MCCS}}(N_1)$ by \eqref{equ:prop_1}.
\EndFor
\State Compute $N_1^{*}=\textrm{argmin}_{N_1\in\Nc}\bar R^{\text{\tiny PF-SA}}_{\text{MCCS}}(N_1)$.
\State Compute $\qbf^*$ by~\eqref{equ:placement} and $\bar R^{\text{\tiny PF-SA}}_{\text{MCCS}}(N_1^*)$ by~\eqref{equ:prop_1}.
\end{algorithmic}
\end{algorithm}

By Proposition~\ref{pro:upper}, for the proposed PF-SA cache placement strategy using two file groups, the optimal $N_1$ for group partition to achieve the minimum average rate can be obtained through a search in $\Nc$. We summarize our proposed algorithm in Algorithm \ref{Alg:Approx_solu}. The algorithm only involves computing the average rate for $N$ times to determine the optimal $N_1$. For each $N_1\in\Nc$, the average rate is computed directly using the expressions in~\eqref{equ:prop_1} and \eqref{equ:two_groups_rate}. Thus, the computational complexity of  Algorithm \ref{Alg:Approx_solu} is much lower than that of the successive GP method in Algorithm~\ref{alg:MCCS1}, which requires successively solving the large-scale GP subproblems. 
Interestingly, our numerical studies in Section~\ref{sec:simu}  show that the average rate achieved by Algorithm~\ref{Alg:Approx_solu} is very close to that by Algorithm~\ref{alg:MCCS1}, and in many cases, it is even lower than that by Algorithm~\ref{alg:MCCS1}.

\subsubsection{Approximate Solutions for Special Scenarios}

We now consider the proposed PF-SA cache placement  in the special scenario of nonuniform file popularity only or nonuniform file size only. Both scenarios have been widely studied in the existing works~\cite{Deng&Dong:TIT22,Daniel&Yu:TIT19,Jin&Cui:Arxiv2018,Niesen&Maddah-Ali:TIT2017,Ji&Order:TIT17,Zhang&Coded:TIT18,Zhang&Lin19Closing,Zhang&Lin15Coded,Cheng&Li17Optimal}.\vspace{.5em}\\
\noindent{\it Nonuniform file popularity only:}
In this case, each file has the same size. Let $F\triangleq F_n, n\in\Nc$. Then, the PF-SA cache placement in~\eqref{equ:placement} becomes
\begin{align}\label{equ:placement_cor}
q_n=
\begin{cases}
\frac{M}{N_1F}, &\quad n\in \Nc_1,\\
0,&\quad n\in \Nc_2.
\end{cases}
\end{align}
Based on~\eqref{equ:placement_cor}, the files in the first group have the same number of bits cached at the users, \ie $q_nF=\frac{M}{N_1}$ for all $n\in\Nc_1$; and the files in the second group $\Nc_2$ are uncached.  Thus,  the decentralized PF-SA cache placement strategy prioritizes popular files and equally allocates the entire user cache to those popular files in the first group. Note that in this case, the subfile size $|W_{n,\Sc}|$ in~\eqref{equ:de_subfile} becomes approximately the same for all $n$'s for user subset $\Sc$. Consequently, the subfiles in a coded message in~\eqref{equ:de_msglength} are of equal size. As a result, the expression of the average rate of the D-MCCS  in~\eqref{equ:two_groups_rate} can be simplified, as shown in the following corollary.

\begin{corollary}\label{cor:upper}
Consider the decentralized caching problem described in Proposition~\ref{pro:upper} with uniform file size $F\triangleq F_1=\ldots= F_N$. The average rate of the D-MCCS  under the PF-SA cache placement is given in \eqref{equ_min_two_group} and \eqref{equ:prop_1} with $R^{\text{\tiny PF-SA}}_{\text{MCCS}}(\dbf_\Ac;N_1)$ expressed as
\begin{align}\label{equ:pupo_two_groups_rate}
R^{\text{\tiny PF-SA}}_{\text{MCCS}}(&\dbf_\Ac;N_1)=\nn\\
&\sum_{s=0}^{A-1}\left( \sum_{i=1}^{\widetilde N(\dbf_\Ac)}\binom{A-i}{s}-\sum_{i=1}^{\widetilde N(\dbf_{\Ac_2})}\binom{A_2-i}{s}  \right)\nn\\
&\ \cdot\left( \frac{M}{N_1F}\right)^{s}\left(1- \frac{M}{N_1F}\right)^{A-s}\!\!F+\widetilde N(\dbf_{\Ac_2})F.
\end{align}
\end{corollary}
\IEEEproof
See Appendix~\ref{proof_cor:upper}.
\endIEEEproof

Following Corollary~\ref{cor:upper}, the optimal $N_1\in\Nc$ that leads to the minimum average rate in this case can be again obtained through a search in $\Nc$, as described in Algorithm~\ref{Alg:Approx_solu}. The only difference is that for given $\dbf_{\Ac}$ and $N_1$, the delivery rate $R^{\text{\tiny PF-SA}}_{\text{MCCS}}(\dbf_\Ac;N_1)$ is computed by~\eqref{equ:pupo_two_groups_rate} instead of~\eqref{equ:two_groups_rate}.  Note in~\eqref{equ:two_groups_rate},  the number of $\max$ operations involved increases exponentially with $A$. In contrast, \eqref{equ:pupo_two_groups_rate} contains only a total of $A(\widetilde N(\dbf_\Ac)+\widetilde N(\dbf_{\Ac_2}))+1$ summation terms, which has a much lower computational complexity than~\eqref{equ:two_groups_rate}.

\begin{Remark}\label{remark3}
For the nonuniform file popularity only scenario, the PF-SA placement is reduced to a two-file-group-based file placement structure similar to the ones considered by~\cite{Ji&Order:TIT17,Zhang&Coded:TIT18} for the D-CCS, where the size of the first group $N_1$ has been proposed through heuristics. However, our work is different from these in the following aspects: First, our placement solution is developed assuming unknown active user set $\Ac$,  ~\cite{Ji&Order:TIT17,Zhang&Coded:TIT18} depend on a known active user set $\Ac$.  Second, the D-MCCS is different from the D-CCS considered in\cite{Ji&Order:TIT17,Zhang&Coded:TIT18} in  the delivery procedure. Specifically, as mentioned in Section~\ref{sec:delivery}, in the D-MCCS, only the coded message for non-redundant groups are delivered, thus removing the redundancy in the coded messages of the D-CCS. Furthermore, for the coded delivery,~\cite{Ji&Order:TIT17,Zhang&Coded:TIT18} apply a user-group-based message generation method, where instead of~\eqref{equ:de_codedmsg}, each coded message is formed by only those files within the same file group, and thus there is no coding across file groups. In contrast, we explore the coded caching gain among all the requested files $\dbf_\Ac$ in Algorithm~\ref{Alg:D-MCCS} using~\eqref{equ:de_codedmsg}. Indeed, it has been shown in the study of the D-CCS that the average rate of the coded delivery that explores the coded caching gain among all files is a lower bound to that of the user-group-based delivery~\cite{Saberali&Lampe:TIT20}.
\end{Remark}

\noindent{\it Nonuniform file size only:}  In this case, the files have the same popularity $p_1=\ldots=p_N$. According to how we index the files as described in  Section~\ref{sec:model}, the files are indexed in decreasing order of their sizes. As a result, under the PF-SA cache placement strategy, the first group $\Nc_1$ contains the $N_1$ largest files, and the cache placement vector $\qbf$ for these files are given in~\eqref{equ:placement}.  The average rate  $\bar R^{\text{\tiny PF-SA}}_{\text{MCCS}}(N_1)$ is still given by \eqref{equ:prop_1} and \eqref{equ:two_groups_rate} in Proposition~\ref{pro:upper}.  Following this, we  can search for the optimal $N_1^*\in\Nc$ using Algorithm~\ref{Alg:Approx_solu} to obtain the minimum average rate under the PF-SA cache placement. Note that different from  the previous scenario, for the nonuniform file size scenario, within the first group $\Nc_1$, the number of cached bits from each file   depends on the file size. As shown in  \eqref{equ:two_groups_rate}, the coded message size depends on $\max_{k\in\Sc}F_{d_k}$. The rate $ R_\text{MCCS}^\text{\tiny PF-SA}(\dbf_{\Ac};N_1)$ 
cannot be further simplified, as we need to determine the largest file size among files requested by each user subset $\Sc$.  Thus, the complexity involved in finding the optimal $N_1^*$ is higher than that in the nonuniform file popularity scenario.  In Section~\ref{sec:sim_size}, through our numerical study, we show that the average rate for the D-MCCS obtained by the optimal PF-SA cache placemen remains to be  very close to that obtained by the successive GP approximation algorithm and the lower bound.

%%%%%%%%%%%%%%%%%%%%%%%%%%%%%%%%%%%%%%%%%%%%%%%%%%%%%%%%%%%%%%%%%%%%%%%%%%%%%%%%
\section{Memory-Rate Tradeoff for Decentralized Caching}\label{sec:lb}
In this section, we characterize the memory-rate tradeoff for decentralized caching under nonuniform file popularity and sizes by proposing a lower bound and comparing it with the average rate of the optimized D-MCCS in {\bf P0}.

%%%%%%%%%%%%%%%%%%%%%%%%%%%%%%%%%%%%%%%%%%%%%%%%%%%%%%%%%%%%%%%%%%%%%%%%%%%%%%%%
\subsection{Lower Bound for Decentralized Caching}
The general idea for developing the lower bound for decentralized caching is to first divide all the possible file demand vectors into different types and then derive a lower bound for each type separately~\cite{Yu&Maddah-Ali:TIT2018}.
Given any active user set $\Ac\subseteq\Kc$, we categorize all the possible demand vectors $\dbf_\Ac\in\Nc^A$ based on the distinct file requests in $\dbf_\Ac$. We denote $\Dc_{\Ac}\triangleq \text{Unique}(\dbf_{\Ac})$ as the set of distinct file indices in demand vector $\dbf_\Ac$, where $\text{Unique}(\dbf_{\Ac})$ is to extract the distinct file indices in $\dbf_\Ac$.
Recall that the leader group $\Uc_{\Ac}$ contains $\widetilde{N}(\dbf_{\Ac})$ users requesting all the distinct files in $\dbf_{\Ac}$. Thus we have $|\Dc_{\Ac}|=|\Uc_{\Ac}|=\widetilde{N}(\dbf_{\Ac})$.
We present a lower bound on the average rate for decentralized caching under nonuniform file popularity and sizes in the following theorem.

\begin{theorem}\label{thm_bnd}
Consider the decentralized caching problem of $N$ files with popularity distribution $\pbf$ and sizes $\{F_i\}$, and $K$ users each with cache size $M$ bits and with probability $p_{\text{a},k}$ being active.  The following optimization problem  provides a lower bound on the average  rate:
\begin{align}
\!\!\!\textrm{\bf P3}\!: \min_{\qbf}&
                    \bar R_{\text{lb}} (\qbf)\triangleq\! E_{\Ac}\bigg[\!\sum_{{\Dc_\Ac}\subseteq\Nc} \sum_{\dbf_\Ac\!\in\Tc\!(\!{\Dc_\Ac}\!)}\!\!\!\bigg(\prod_{k\in\Ac}\!p_{d_k}\!\bigg)\! R_\text{lb}({\Dc_\Ac};\!\qbf)\bigg]\label{equ:lb_obj1} \\
      \textrm{s.t.}&\ \eqref{equ:de_cons2},\eqref{equ:de_cons1}\nn
      \end{align}
where $\Tc({\Dc_\Ac})\triangleq \{\dbf_\Ac: \text{Unique}(\dbf_\Ac)=\Dc_\Ac, \ \dbf_\Ac\in \Nc^A\}$, and $R_\text{lb}({\Dc_\Ac};\qbf)$ is the  lower bound on the rate for given $\qbf$ and ${\Dc}_\Ac$, given by
\begin{align}\label{R_lblb}
\!\!\!&R_\text{lb}(\Dc_\Ac;\qbf)\triangleq\nn\\
\!\!\!\!&\max_{\pi:\Ic_{|\Dc_\Ac|}\rightarrow\Dc_\Ac}\! \sum_{s=0}^{A-1}\sum_{i=1}^{\widetilde{N}(\dbf_\Ac)}\!\!\!\binom{A-i}{s}q_{\pi(i)}^{s}(1\!-\!q_{\pi(i)})^{A-s}F_{\pi(i)}
\end{align} where $\Ic_{|\Dc_\Ac|}\triangleq\{1,\ldots,|\Dc_\Ac|\}$, and  $\pi:\Ic_{|\Dc_\Ac|}\to\Dc_\Ac$ is any bijective map from $\Ic_{|\Dc_\Ac|}$ to $\Dc_\Ac$.
\end{theorem}
\IEEEproof
See Appendix \ref{proof_thm_bnd}
\endIEEEproof

Note that {\bf P3} is a non-convex optimization problem, and the only difference between {\bf P3} and {\bf P0} are their objective functions $\bar R_{\text{MCCS}}(\qbf)$ and $\bar R_\text{lb}(\qbf)$.
Thus, we can solve {\bf P3} using an approach similar to Algorithm~\ref{alg:MCCS1} for {\bf P0} in Section~\ref{sec:SP_MCCS}. We first formulate {\bf P3} into an equivalent CGP problem. To do so, with the same auxiliary variables $x_n, n\in\Nc$, we add the inequality constraints
\eqref{equ:de_cons3}.
Also, we introduce  auxiliary variable $r_{\Dc_{\Ac}}$ for $\Dc_{\Ac}\subseteq\Nc$ and  $\Ac\subseteq\Kc$. Similar to~\eqref{equ:de_cons4}, based on~\eqref{equ:de_cons3}, we replace the expression in \eqref{R_lblb} by $r_{\Dc_{\Ac}}$ and add the following constraints
\begin{align}\label{equ:lb_cons_4_0}
\sum_{s=0}^{A-1}\sum_{i=1}^{\widetilde N(\!\dbf_\Ac\!)}\binom{A-i}{s}\!\left(q_{\pi(i)}\right)^{s}\!&\left(x_{\pi(i)}\right)^{A-s}F_{\pi(i)}\!\leq\! r_{\Dc_\Ac},\nn\\
&\hspace{4em}\forall\pi:\Ic_{|\Dc_\Ac|}\rightarrow\Dc_\Ac
\end{align}
for given $\Dc_{\Ac}\subseteq\Nc$ and  $\Ac\subseteq\Kc$.
Similar to the reformulation of {\bf P0} to {\bf P1}, with \eqref{equ:lb_cons_4_0}, we can reformulate {\bf P3} into the following CGP.
\begin{align}
&\textrm{\bf P4}: \;\min_{\qbf\succcurlyeq0,\xbf\succcurlyeq0,\rbf\succcurlyeq0}E_{\Ac}\bigg[ \sum_{\Dc_{\Ac}\subseteq\Nc}\sum_{\dbf_\Ac\in\Tc({\Dc_\Ac})}\bigg(\prod_{k\in\Ac}p_{d_{k}}\bigg)r_{\Dc_\Ac}\bigg] \nn\\
&\textrm{s.t.} \;\; \eqref{equ:MCCS_cons_1},\eqref{equ:MCCS_cons_2},\eqref{equ:MCCS_cons_3}\ \text{and}\nn\\
 & (r_{\Dc_\Ac})^{-1}\sum_{s=0}^{A-1}\sum_{i=1}^{\widetilde N(\dbf_\Ac)}\!\!\binom{A-i}{s}\!\left(q_{\pi(i)}\right)^{s}\!\!\left(x_{\pi(i)}\right)^{A-s}\!\!F_{\pi(i)} \leq 1,\nn\\
 &\quad\quad\quad\quad\quad\quad\quad\Ac\subseteq\Kc, \Dc_{\Ac}\subseteq\Nc, \forall\pi:\Ic_{|\Dc_\Ac|}\rightarrow\Dc_\Ac. \label{equ:lb_cons_4}
\end{align}

Following the similar steps in Section~\ref{sec:SP_MCCS}, we use the successive GP approximation algorithm to solve {\bf\ P4}. Let $\bar R^\text{\tiny CGP}_\text{lb}(\qbf,\xbf,\rbf)$  denote the objective function of {\bf P4}. Let $(\qbf^{(i)},\xbf^{(i)},\rbf^{(i)})$ denote the solution obtained in iteration $i$. In iteration $i+1$, given $(\qbf^{(i)},\xbf^{(i)})$, we formulate the following approximate optimization problem of {\bf P4}:
\begin{align}
\textrm{\bf P5}\big(\qbf^{(i)},\xbf^{(i)}\big): &\min_{\qbf\succcurlyeq0,\xbf\succcurlyeq0,\rbf\succcurlyeq0}\bar R^\text{\tiny CGP}_\text{lb}(\qbf,\xbf,\rbf)  \nn\\
&\ \ \textrm{s.t.} \;\; \eqref{equ:MCCS_cons_1},\eqref{equ:MCCS_cons_2},\eqref{equ:MCCS_cons_5},\ \text{and}\ \eqref{equ:lb_cons_4}.\nn
\end{align}

We then iteratively solve \textrm{\bf P5}$\left(\qbf^{(i)},\xbf^{(i)}\right)$ to obtain a stationary point of {\bf P4}. 
Finally, by the equivalence of {\bf P3} and {\bf P4}, we obtain the stationary point of {\bf P3}. The algorithm is summarized in Algorithm
\ref{alg:Bnd}. %Note that the size of the {\bf P5} we solve at each iteration is $N!K$, which is more manageable than that of Algorithm~\ref{alg:MCCS1}.

\begin{algorithm}[t]
\caption{The successive GP approximation algorithm for {\bf P3}}\label{alg:Bnd}
\begin{algorithmic}[1]
\Require
   { $K$,  $M$,  $N$, $\pbf$, $\pbf_{\text{a}}$ }
\Ensure {$\bar R_{\text{lb}}^{*}$, $\qbf^*$}\\
{\bf Initialization: }Choose initial feasible point $\left(\qbf^{(0)},\xbf^{(0)}, \rbf^{(0)}\right)$. set $i=0$.
\Repeat
    \State Solve {\bf P5}$\left(\qbf^{(i)},\xbf^{(i)}\right)$ to obtain $\left(\qbf^{(i+1)},\xbf^{(i+1)}, \rbf^{(i+1)}\right)$.
    \State Set $i=i+1$.
\Until $\bar R^\text{\tiny CGP}_{\text{lb}}(\qbf^{(i)},\xbf^{(i)},\rbf^{(i)})$ converges.\\
Set $\bar R_{\text{lb}}^{*}=\bar R^\text{\tiny CGP}_{\text{lb}}(\qbf^{(i)},\xbf^{(i)},\rbf^{(i)})$; $\qbf^*=\qbf^{(i)}$.
%  at $\bar R_{\text{lb}}^{(i)}-\bar R_{\text{lb}}^{(i+1)}\leq0.000001$
\end{algorithmic}
\end{algorithm}

%%%%%%%%%%%%%%%%%%%%%%%%%%%%%%%%%%%%%%%%%%%%%%%%%%%%%%%%%%%%%%%%%%%%%%%%
\subsection{Memory-Rate Tradeoff Characterization}\label{sec:tradeoff}
We now compare the optimized D-MCCS in {\bf P0} with the lower bound in {\bf P3} and demonstrate the equivalence of the two problems in some special cases.
Since the difference between {\bf P0} and {\bf P3} is only in the expression of average rate objective function, it is sufficient to compare $\bar R_\text{MCCS}(\qbf)$ and $\bar R_\text{lb}(\qbf)$.

We first consider a special case where  there are at most two active users at the same time, \ie $A \le 2$. Conditioned on $A\le2$, we rewrite $\bar R_\text{MCCS}(\qbf)$ in \eqref{equ:de_obj1} and $\bar R_\text{lb}(\qbf)$ in \eqref{equ:lb_obj1} as
\begin{align}
&\!\!\!\bar R_\text{MCCS}(\qbf)\!=\!E_{\Ac}\bigg[\sum_{\dbf_{\Ac}\!\in\Nc^{A}}\!\!\!\bigg(\prod_{k\in\Ac}\!\!p_{d_{k}}\!\!\bigg)\!R_{\text{MCCS}}(\dbf_\Ac;\!\qbf) \Big{|} A\!\le\!2\bigg],\label{A2_MCCS}\\
%\end{align}
%\begin{align}
&\!\!\!\bar R_{\text{lb}} (\qbf)\!=\! E_{\Ac}\!\bigg[\!\sum_{{\Dc_\Ac}\subseteq\Nc} \sum_{\dbf_\Ac\!\in\Tc\!(\!{\Dc_\Ac}\!)}\!\!\!\!\bigg(\!\prod_{k\in\Ac}\!\!p_{d_k}\!\!\bigg) R_\text{lb}({\Dc_\Ac};\!\qbf)\Big{|}A\!\le\!2\bigg].\!\! \label{A2_lb}
\end{align}
Comparing \eqref{A2_MCCS} and \eqref{A2_lb}, we show in the following theorem that the lower bound in {\bf P3} is tight.
\begin{theorem}\label{thm_K2}
Assume there are no more than two actives users at the same time, \ie $A\le2$. The average rate of the optimized D-MCCS in {\bf P0} attains the lower bound in {\bf P3}.
\end{theorem}
\IEEEproof
See Appendix~\ref{proof_thm_K2}.
\endIEEEproof

Theorem~\ref{thm_K2} shows that if there are no more than two active users at the same time, then the optimized  D-MCCS is  an optimal   decentralized caching scheme. In this case, the optimized D-MCCS characterizes the exact memory-rate tradeoff for decentralized caching under nonuniform file popularity and sizes. Recall that zero-padding is used in the coded delivery phase of the D-MCCS. Theorem~\ref{thm_K2} implies that the use of zero-padding incurs no loss of optimality for $A\le2$ active users.
{Specifically, we note that zero-padding is only applied when there are $A=2$ active users. For the case of $A=1$ active user,
there is only one file being requested. In this case, the coded message in~\eqref{equ:de_codedmsg} only contains one subfile, and no zero-padding is needed. }

% In the general scenario where the active user set  is not limited to two users, our simulation results in Section~\ref{} show that the gap is in general very small between the average rate of the D-MCCS and the lower bound {\bf P3}.

{When $A>2$, in general, it is difficult to establish the equivalency of {\bf P0} and {\bf P3} because of the difference between $\bar R_\text{MCCS}(\qbf)$ and $\bar R_\text{lb}(\qbf)$. However, when all files are of the same size but only different in popularity,  we show that {\bf P0} and {\bf P3} can still be the same under a certain condition, as described in the following proposition.}

\begin{proposition}\label{prop_1}
For decentralized caching with nonuniform file popularity and uniform file size $F\triangleq F_1=\ldots=F_N$, if $\qbf^*$ with $q^*_1=\cdots=q^*_N$ is the optimal solution to both {\bf P0} and {\bf P3}, then $\bar R_{\text{MCCS}}(\qbf^*)=\bar R_{\text{lb}}(\qbf^*)$, and the optimized D-MCCS in {\bf P0} attains the lower bound in {\bf P3}.
\end{proposition}
\IEEEproof
See Appendix~\ref{proof_prop_2}.
\endIEEEproof
Proposition~\ref{prop_1} indicates that for files with the same size but only popularity may be different, if the optimal placement $\qbf^*$ is symmetric for all files, $q^*_1=\cdots=q^*_N$, then the optimized D-MCCS is an optimal decentralized caching scheme, which characterizes the exact memory-rate tradeoff for decentralized caching. One known example satisfying the condition is the special case of uniform file popularity and size. In this case, the optimized D-MCCS ({\bf P0}) and the lower bound ({\bf P3}) have the same optimal solution $\qbf^*$ with $q^*_n$'s  being  all identical. In this case, it has been shown in~\cite{Yu&Maddah-Ali:TIT2018} that $\bar{R}_{\text{MCCS}}(\qbf^*)=\bar{R}_{\text{lb}}(\qbf^*)$. 

Finally, we point out that, even though {\bf P0} does not attain {\bf P3} in general, our numerical study in Section \ref{sec:simu} shows that the gap between the optimized D-MCCS and the lower bound in {\bf P3} is typically very small.
This indicates that the performance of the optimized D-MCCS is very close to the optimal decentralized caching.

%%%%%%%%%%%%%%%%%%%%%%%%%%%%%%%%%%%%%%%%%%
\section{Simulation Results}\label{sec:simu}

In this section, we evaluate the performance of the D-MCCS in {\bf P0} and compare it with the proposed lower bound for decentralized caching in {\bf P3}. For solving {\bf P0}, we consider the proposed successive GP approximation method in Algorithms~\ref{alg:MCCS1} and the PF-SA-cache-placement-based scheme in Algorithm~\ref{Alg:Approx_solu}. For solving {\bf P3}, we use the successive GP approximation method in Algorithm~\ref{alg:Bnd}.  We use $\bar R$ to denote the average rate obtained by various schemes and the lower bound considered in the simulation.
For both Algorithms~\ref{alg:MCCS1} and~\ref{alg:Bnd}, we set the convergence criterion to be that the difference in the average rate $\bar R$ over two consecutive iterations is less than $10^{-4}$. In our simulation, we set the probability of each user being active  as $p_{\text{a},k}=0.5$, for $k\in\Kc$.

\renewcommand{\arraystretch}{1}
\begin{table}[t]
\centering
\caption{The list of file populariti and sizes, for $N=6$, $8$ and $10$ files.}
\resizebox{8.65cm}{!}{\begin{tabular}{|cc|cc|cc|}
\hline
\multicolumn{2}{|c|}{$N=6$}             & \multicolumn{2}{c|}{$N=8$}            & \multicolumn{2}{c|}{$N=10$}         \\ \hline
\multicolumn{1}{|c|}{Popularity} & \multicolumn{1}{c|}{Size (kbit)} & \multicolumn{1}{c|}{Popularity} & Size (kbit) & \multicolumn{1}{c|}{Popularity} & Size (kbit) \\ \hline
\multicolumn{1}{|c|}{0.4643} & 0.1667 & \multicolumn{1}{c|}{0.4286} & 0.625 & \multicolumn{1}{c|}{0.4052} & 0.1 \\ \hline
\multicolumn{1}{|c|}{0.2021} & 0.3333 & \multicolumn{1}{c|}{0.1866} & 0.125 & \multicolumn{1}{c|}{0.1764} & 0.2 \\ \hline
\multicolumn{1}{|c|}{0.1242} & 0.5    & \multicolumn{1}{c|}{0.1147} & 0.25  & \multicolumn{1}{c|}{0.1084} & 0.3 \\ \hline
\multicolumn{1}{|c|}{0.088}  & 0.8333 & \multicolumn{1}{c|}{0.0812} & 0.875 & \multicolumn{1}{c|}{0.0786} & 0.4 \\ \hline
\multicolumn{1}{|c|}{0.0673} & 1      & \multicolumn{1}{c|}{0.0621} & 0.5   & \multicolumn{1}{c|}{0.0587} & 0.5 \\ \hline
\multicolumn{1}{|c|}{0.0541} & 0.6667 & \multicolumn{1}{c|}{0.0499} & 0.375 & \multicolumn{1}{c|}{0.0472} & 0.9 \\ \hline
\multicolumn{1}{|l|}{}       &        & \multicolumn{1}{c|}{0.0415} & 0.75  & \multicolumn{1}{c|}{0.0392} & 0.6 \\ \hline
\multicolumn{1}{|l|}{}       &        & \multicolumn{1}{c|}{0.0353} & 1     & \multicolumn{1}{c|}{0.0334} & 0.7 \\ \hline
\multicolumn{1}{|l|}{}       &        & \multicolumn{1}{c|}{}       &       & \multicolumn{1}{c|}{0.029}  & 0.8 \\ \hline
\multicolumn{1}{|l|}{}       &        & \multicolumn{1}{c|}{}       &       & \multicolumn{1}{c|}{0.0256} & 1   \\ \hline
\end{tabular}}\label{tab:index1}\vspace*{-1.5em}
\end{table}

\begin{figure}[t]
  \centering
  \includegraphics[width=0.45\textwidth]{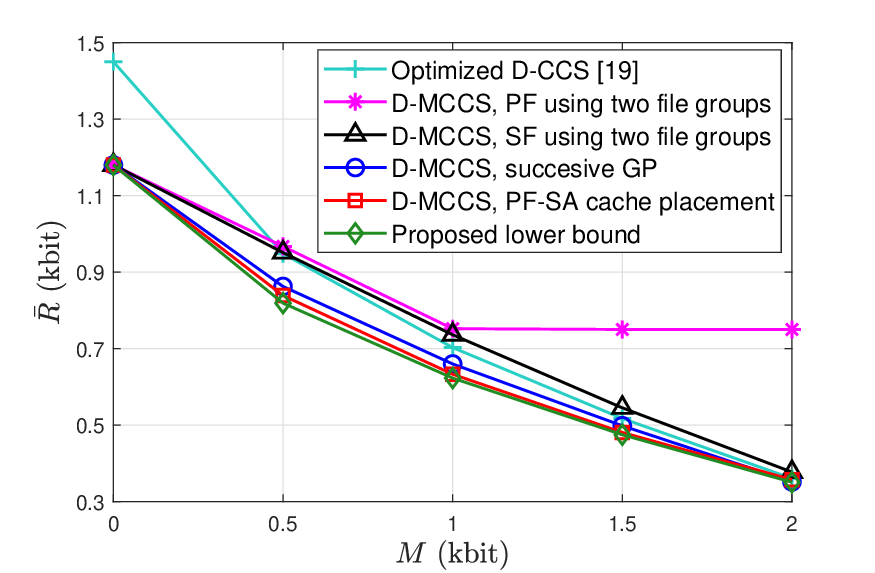}\vspace*{-.5em}
  \caption{Average rate $\bar R$ vs. cache size $M$  ($K=4$, $N=6$. File popularity and size are described in Table~\ref{tab:index1}.).}\label{performance_1}
%\end{figure}
%\begin{figure}[t]
  \centering
  \includegraphics[width=0.45\textwidth]{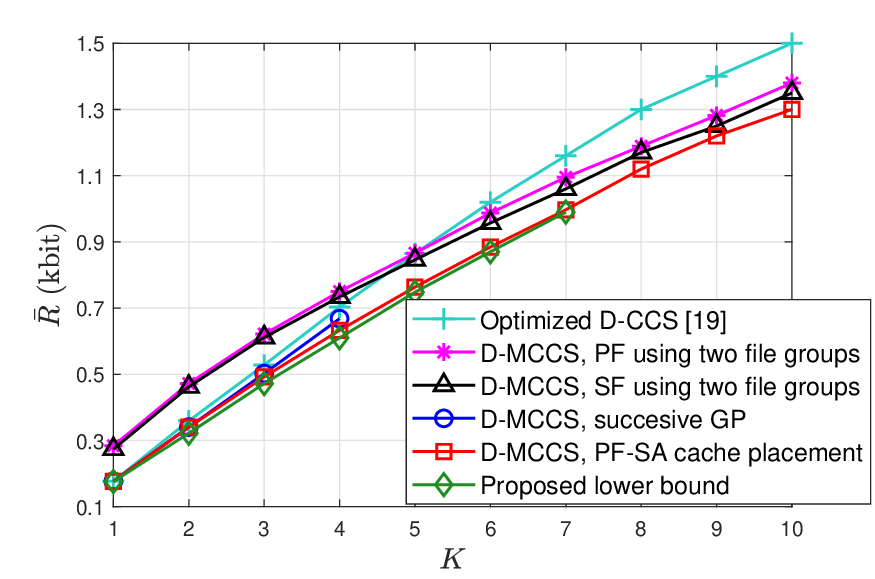}\vspace*{-.5em}
  \caption{Average rate $\bar R$ vs. number of users $K$ ($M=1$ kbit, $N=6$. File popularity and size are described in Table~\ref{tab:index1}.).}\label{performance_2_RvsK}
\vspace*{-1.5em}
\end{figure}
\renewcommand{\arraystretch}{1.1}
\begin{table*}[t]
\centering
\caption{Average Computation time (sec.)  of Proposed Algorithms in Fig.~\ref{performance_1}. }
\vspace{-1em}
\label{table1}
\begin{tabular}{|p{14.5em}<{\centering}|p{2.5em}<{\centering}|p{2.5em}<{\centering}|
p{2.5em}<{\centering}|p{2.5em}<{\centering}|p{2.5em}<{\centering}|p{2.8em}<{\centering}
|p{2.8em}<{\centering}|}
\hline
%\multicolumn{1}{|c|}{}
$M$ (kbit)  &0.5  & 1 &1.5& 2&2.5 & 3 &3.5  \\ \hline%\hline
D-MCCS, successive GP &25,132 &14,103 &21,459  & 30,504&38,349  &42,828  &39,825       \\ \hline
D-MCCS, PF-SA cache placement   &1.8             & 1.8& 1.8&1.8 &1.8 & 1.8  & 1.8      \\ \hline
\end{tabular}\\[-1em]
\end{table*}

\renewcommand{\arraystretch}{1.1}
\begin{table*}[t]
\centering
\caption{Average Computation time (sec.)  of Proposed Algorithms in Fig.~\ref{performance_2_RvsK}. }
\vspace{-1em}
\label{table2}
\begin{tabular}{|p{14.5em}<{\centering}|p{2.5em}<{\centering}|p{2.5em}<{\centering}|
p{2.5em}<{\centering}|p{2.5em}<{\centering}|p{2.5em}<{\centering}|p{2.8em}<{\centering}
|p{2.8em}<{\centering}|p{2.5em}<{\centering}|p{2.5em}<{\centering}|p{2.5em}<{\centering}
|}
\hline
%\multicolumn{1}{|c|}{}
$K$  &1  & 2 &3& 4&5 & 6 &7& 8 & 9 & 10 \\ \hline%\hline
D-MCCS, successive GP &1.2 &8.9 & 521 &38,349 &N/A  &N/A  &N/A  & N/A & N/A   & N/A     \\ \hline
D-MCCS, PF-SA cache placement  &0.03&0.05 &0.19 &2.6 &8.92 &42.36   &69.62   &116.01    &178.92    &258.2    \\ \hline
\end{tabular}\\[-1.2em]
\end{table*}

%%%%%%%%%%%%%%%%%%%%%%%%%%%%%%%%%%%%%%%%%%%%%%%%%%%%%%%%
\subsection{Nonuniform File Popularity and Sizes}\label{sec:simA}

We first consider the case where files have different popularity and sizes.  We list the file popularity distribution and size  for $N=6,8,10$ files in Table~\ref{tab:index1}. They are used in our simulation.  The file popularity distribution is generated using Zipf distribution as $p_{n}={n^{-\theta}}/{\sum_{i=1}^{N}i^{-\theta}}$, where the Zipf parameter $\theta=1.2$. To compare different cache placement strategies, besides the D-MCCS with our proposed two algorithms (Algorithms~\ref{alg:MCCS1} and~\ref{Alg:Approx_solu}) and the proposed lower bound (Algorithm~\ref{alg:Bnd}), we also consider the following methods: \emph{ i) The D-MCCS with the SF strategy using two file groups}: sort files based on their sizes and partition them into two groups with the first group containing the $N_1$ largest files;  set cache allocation for each file $n$ in the first group as  $\min\{M/N_1, F_n\}$; search for the optimal $N_1^*$ that gives the minimum average rate.
\emph{ii) The D-MCCS with the PF strategy using two file groups}: since files are already indexed based on popularity,  partition them into two groups with the first group containing the $N_1$ most popular files; set cache allocation to file $n$ as $\min\{M/N_1, \min\{F_1,\cdots,F_n\}\}$, for $1\le n\le N_1$; search for the optimal $N_1^*$ that gives the minimum average rate.
 \emph{iii) The optimized D-CCS in~\cite{Wang2019Optimization}.}

We consider $N=6$ files in Table~\ref{tab:index1} for  both Figs.~\ref{performance_1} and~\ref{performance_2_RvsK}. In Fig~\ref{performance_1}, we plot the average rate $\bar R$ vs. the cache size $M$   by different methods, for $K=4$ users.
First, we observe that the PF-SA cache placement outperforms both PF and SF strategies with a noticeable performance gap. In particular, note that the average rate $\bar{R}$ of  the D-MCCS with the SF strategy   is  higher than that of the optimized D-CCS for $M\ge0.5$ kbit.  This shows that only prioritizing the file size but ignoring the nonuniform file popularity in the cache placement design results in a worse performance. For the PF strategy that only prioritizes the file popularity, the resulting $\bar R$ is even higher than that of the SF strategy. In particular,  for $M\ge1$ kbit, the average rate of the PF strategy is floored and no longer reduces  despite  $M$ increases. This is due to the drawback of the PF strategy  discussed in Remark~\ref{remark:SFvsPF} that cache memories are not fully utilized in this case, as the  most popular file is of relatively smaller size  0.1667 kbit, and at most around 1 kbit of contents from 6 files can be cached. Between our two proposed algorithms for D-MCCS, the average rate $\bar R$ achieved by the PF-SA-cache-placement-based approach  is always slightly lower than that of the successive GP approximation method, while the former has much lower complexity to implement than the latter. Finally, we observe that the average rate $\bar R$ by the D-MCCS with the PF-SA cache placement is very close to the proposed lower bound. This indicates the effectiveness of our proposed simple PF-SA cache placement strategy and the near-optimal performance of the optimized D-MCCS. 

Fig.~\ref{performance_2_RvsK} shows the average rate $\bar R$ vs. the number of users $K$ for $M=1$ kbit. Due to the high computational complexity of  the successive GP approximation algorithm  in Algorithm~\ref{alg:MCCS1} as $K$ increases, we only show its result for $K \le 4$, and similarly the result of Algorithm~\ref{alg:Bnd} for the lower bound for $K \le 7$. Again, the D-MCCS with PF-SA cache placement outperforms all the methods considered. It achieves the lowest $\bar R$ that is very close to the lower bound. This again demonstrates the near-optimal performance of the PF-SA cache placement strategy and the optimized D-MCCS. The performance gap between   our proposed algorithms  and the optimized D-CCS increases with $K$. The reason is that there are more redundant file requests as the number of users increases; As a result, the D-CCS produces more redundant messages  for delivery, while these redundant messages are eliminated by the D-MCCS. %We also see the   gap between the PF-SA cache placement and the SF strategy  reduces as $K$ increases.  

Table~\ref{table1} shows the average computation time of the successive GP approximation method in Algorithm~\ref{alg:MCCS1} and the PF-SA-cache-placement-based approach in Algorithm~\ref{Alg:Approx_solu}  in generating  Fig.~\ref{performance_1}.
We have used MATLAB 2021b on a Windows x64 machine equipped with Intel 11th i5 CPU with 4.6 GHz and 32 GB RAM. The computation time of the  PF-SA-cache-placement-based approach is significantly lower than that of the successive GP method, and it remains unchanged for different values of $M$. Similarly, Table~\ref{table2} shows the average computation time of the two algorithms to generate Fig.~\ref{performance_2_RvsK} for different values of $K$. We see that the computational complexity  of the successive GP approximation method increases very fast with $K$ and becomes impractical for $K>4$ users. In contrast,  the computational complexity for the simple PF-SA-cache-placement-based approach  increases only mildly  with $K$ at a much slower growth rate.
\begin{figure}[t]
  \centering
  \includegraphics[width=0.45\textwidth]{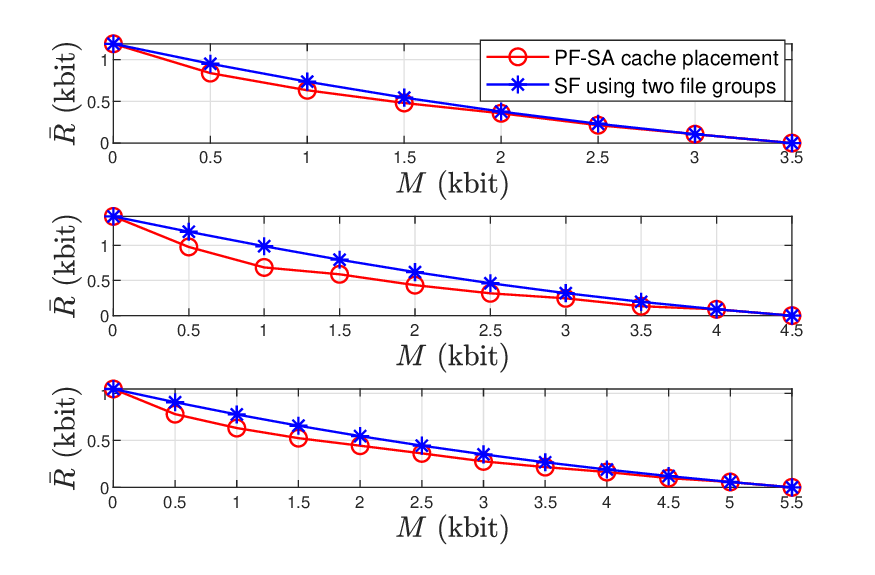}\vspace*{-1em}
  \caption{Comparison of the PF-SA cache placement strategy and the SF using two file groups cache placement strategy ($K=5$).  Top: $N=6$, middle: $N=8$, bottom: $N=10$. File popularity and size for each case are shown in Table~\ref{tab:index1}. }\label{fig:popu_vs_length_v1}\vspace*{-1em}
\end{figure}

For different sets of files in Table~\ref{tab:index1}, we further compare  the proposed PF-SA cache placement and the SF strategy using two file groups. In Fig.~\ref{fig:popu_vs_length_v1}, we plot the average rate $\bar R$ vs. the cache size $M$ under these two cache placement strategies for $K=5$ users and $N=6,8,10$ files. The proposed PF-SA cache placement always achieves a lower value of $\bar R$ than the SF strategy, for all values of $M$ considered. The performance gap of the two cache placement strategies is more noticeable for a small to moderate cache size $M$, indicating that file popularity is more critical than file size in designing a cache placement strategy when the cache storage is limited. 
%This indicates that the nonuniform file popularity and sizes needs to be addressed more carefully for a small $M$. 
In general, Fig.~\ref{fig:popu_vs_length_v1} verifies the advantage of the proposed PF-SA cache placement, where  the nonuniformity in both the file popularity and size is considered for the cache placement design for the D-MCCS.

%%%%%%%%%%%%%%%%%%%%%%%%%%%%%%%%%%%%%%%%%%%%%%%%%%%%%%%%
\subsection{Nonuniform File Popularity Only}\label{sec:simB}

\begin{figure}[t]
 \centering
  \includegraphics[width=0.45\textwidth ]{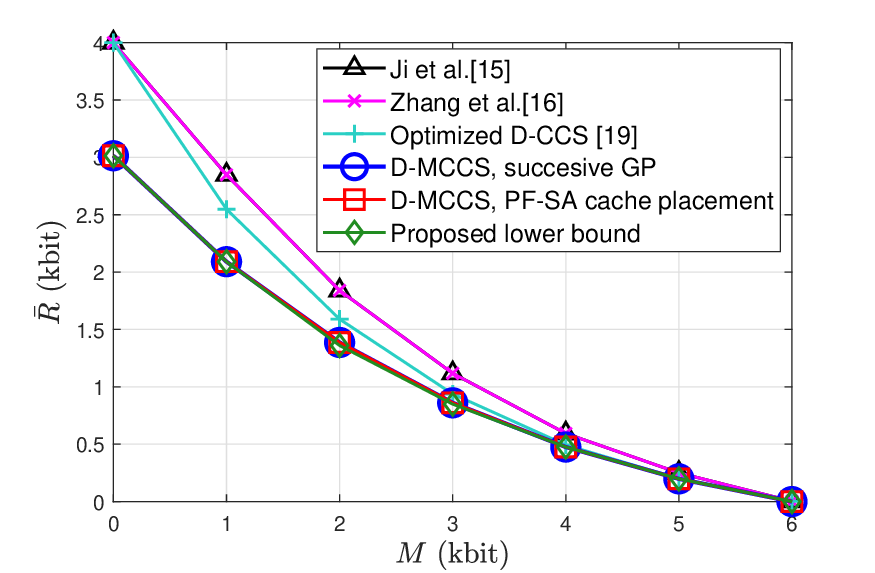}\vspace*{-.5em}
  \caption{Average rate $\bar{R}$ vs. cache size $M$  ($K=4$, $N=6$,  Zipf file popularity distribution with $\theta=0.56$, equal file size.).}
  \label{fig:R_vs_M_56}
  \end{figure}

We now assume all files have the same size and focus on the nonuniform file popularity only scenario. In this case, we can compare our schemes with several existing D-CCS based decentralized caching schemes, which are designed in this scenario.
In particular, we will consider the  well-known  D-CCS based decentralized caching schemes in~\cite{Ji&Order:TIT17} and~\cite{Zhang&Coded:TIT18} and the optimized D-CCS in~\cite{Wang2019Optimization}. In Fig. \ref{fig:R_vs_M_56}, we plot the average rate $\bar R$ vs. cache size $M$, for $N=6$ files and $K=4$ users. For the nonuniform file popularity distribution, we set the Zipf parameter $\theta=0.56$.
For the D-MCCS in {\bf P0}, we observe that the successive GP approximation method and the PF-SA-cache-placement-based approach achieve nearly identical performance. Among the caching schemes considered, the D-MCCS provides the lowest $\bar R$ for all values of  $M$. We observe that the gap between the D-MCCS-based schemes and the D-CCS-based schemes is bigger as $M$ becomes smaller. This is because that for a smaller cache size $M$, there typically exist more redundant messages, and the D-MCCS can reduce this redundancy in coded delivery to achieve a larger coded caching gain. Especially for $M=0$, when all requested files need to be delivered, the amount of such redundancy is the largest.
The average rate $\bar R$ by the D-MCCS obtained through the successive GP approximation method and the PF-SA-cache-placement-based approach are both very close to the lower bound in {\bf P3}. There is only a very small gap observed for $M=2$ or $3$ kbit. To see the gap clearly, the values of the average rate by these methods for $M=2$ and $3$ kbit  are shown in Table~\ref{M23}. The small difference again demonstrates the near-optimal performance of the optimized D-MCCS under the nonuniform file popularity.
\begin{table}[t]
\renewcommand{\arraystretch}{1.1}
\centering
\caption{Average rate $\bar R$ (kbit) of Proposed Algorithms in Fig.~\ref{fig:R_vs_M_56}.}
\vspace{-.5em}
\label{M23}
\begin{tabular}{|p{15.2em}<{\centering}|p{2.5em}<{\centering}|p{2.5em}<{\centering}|}
\hline
%\multicolumn{1}{|c|}{}
$M$ (kbit)  &2  & 3  \\ \hline%\hline
D-MCCS, successive GP &1.3867 &0.8607\\ \hline
D-MCCS, PF-SA cache placement &1.3867 &0.8607\\ \hline
Lower bound   &1.3601             & 0.8493\\ \hline
\end{tabular}\\[-1.2em]
\end{table}

\begin{figure}[t]
   \centering
  \includegraphics[width=0.45\textwidth]{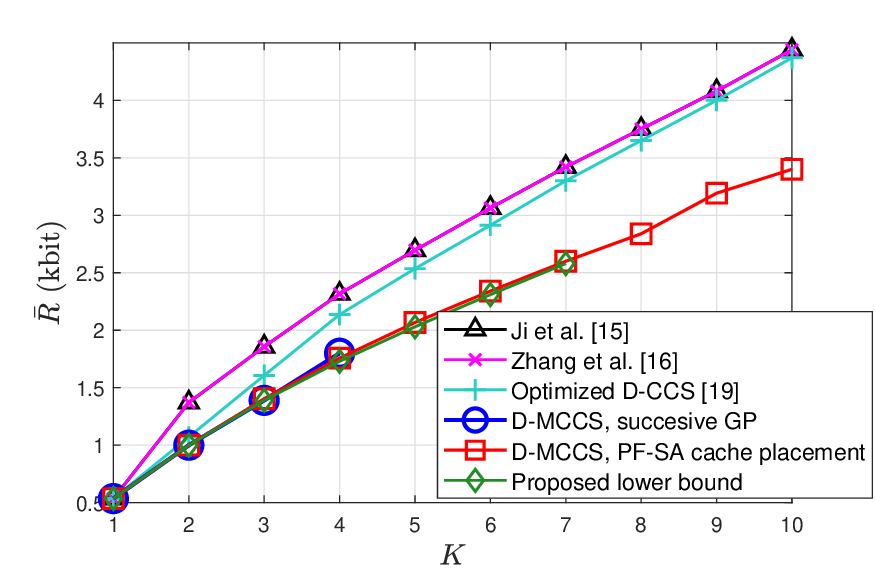}\vspace*{-.5em}
  \caption{Average rate $\bar{R}$ vs. number of users $K$  ($M=1$ kbit, $N=6$, Zipf file popularity distribution with $\theta=1.2$, equal file size.).}
  \label{fig:R_vs_M_1.2}\vspace*{-1em}
\end{figure}

In Fig. \ref{fig:R_vs_M_1.2}, we  plot the average rate $\bar R$ vs. the number of users $K$, for $N=6$ files and $M=1$ kbit. We consider Zipf parameter $\theta=1.2$ for a more diverse file popularity distribution. Among all the schemes compared, the D-MCCS with the PF-SA cache placement  achieves the lowest $\bar R$ for all values of  $K$. The average rates obtained by the successive GP approximation and PF-SA-cache-placement-based algorithms are again nearly identical. In particular, for $K=4$, the achieved $\bar R$ by the PF-SA-cache-placement-based approach is slightly lower than that of  the successive GP approximation method. This again shows that the PF-SA-cache-placement-based algorithm could perform even better than the more computationally complicated successive GP approximation algorithm.  Its gap to the lower bound is very small in general, demonstrating that the optimized D-MCCS is near optimal for decentralized caching.

The average computation time of our two proposed algorithms for the D-MCCS in generating Fig.~\ref{fig:R_vs_M_1.2} is shown in Table~\ref{table-popuOnly}, for different number of users $K$. The PF-SA-cache-placement-based approach is very fast in computing a solution. Moreover, compared with Table~\ref{table2}, we see that when files have the same size but different popularity, the computation time of the PF-SA-cache-placement-based approach grows at a much lower rate as $K$ increases. This is because for the nonuniform file popularity only scenario, the expression of $\bar R$ in~\eqref{equ:pupo_two_groups_rate} is used, which has much less computation complexity than that of~\eqref{equ:two_groups_rate} for nonuniform file popularity and size.

\begin{table*}[t]
\centering
\caption{Average Computation time (sec.)  of Proposed Algorithms in Fig.~\ref{fig:R_vs_M_1.2}. }
\vspace{-0.9em}
\label{table-popuOnly}
\begin{tabular}{|p{14.5em}<{\centering}|p{2.5em}<{\centering}|p{2.5em}<{\centering}|
p{2.5em}<{\centering}|p{2.5em}<{\centering}|p{2.5em}<{\centering}|p{2.8em}<{\centering}
|p{2.8em}<{\centering}|p{2.5em}<{\centering}|p{2.5em}<{\centering}|p{2.5em}<{\centering}
|}
\hline
%\multicolumn{1}{|c|}{}
$K$  &1  & 2 &3& 4&5 & 6 &7& 8 & 9 & 10 \\ \hline%\hline
D-MCCS, successive GP &1.2 &8.9 & 521 &38,349 &N/A  &N/A  &N/A  & N/A & N/A   & N/A     \\ \hline
D-MCCS, PF-SA cache placement  &0.03&0.05 &0.09 &1.2 &1.27 &1.9   &2.06   &2.22    &2.58    &3.2    \\ \hline
\end{tabular}\\[-1em]
\end{table*}

\begin{table*}[t]
\centering
\caption{Average Computation time (sec.)  of Proposed Algorithms in Fig.~\ref{performance_3_RvsN}.}
\vspace{-.9em}
\label{table3}
\begin{tabular}{|p{14.5em}<{\centering}|p{2.5em}<{\centering}|p{2.5em}<{\centering}|
p{2.5em}<{\centering}|p{2.5em}<{\centering}|p{2.5em}<{\centering}|p{2.8em}<{\centering}
|p{2.8em}<{\centering}|p{2.5em}<{\centering}|p{2.5em}<{\centering}|}
\hline
%\multicolumn{1}{|c|}{}
$N$  &2  & 3 &4& 5&6 & 7 &8& 9 & 10  \\ \hline%\hline
D-MCCS, successive GP& 28&2,219&9,188 &19,258 &38,349    &N/A  &N/A  & N/A & N/A    \\ \hline
D-MCCS, PF-SA cache placement  &1.8             & 2& 2.6&3 &3.5 & 4.2  & 5  & 5.8   & 6.6   \\ \hline
\end{tabular}\\[-1em]
\end{table*}

%%%%%%%%%%%%%%%%%%%%%%%%%%%%%%%%%%%%%%%%%%%%%%%%%%%%%%%%
\subsection{Nonuniform File Size Only}\label{sec:sim_size}
\begin{figure}[t]
  \centering
  \includegraphics[width=0.45\textwidth]{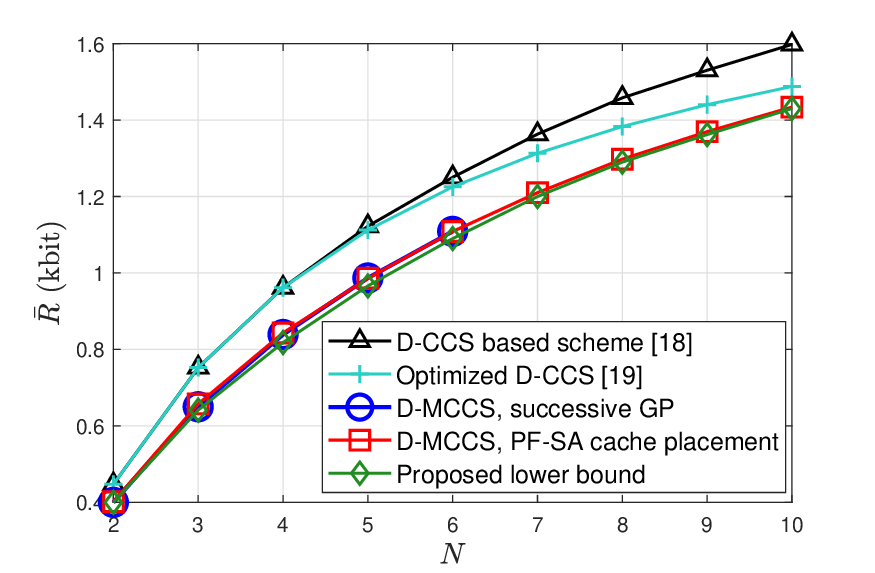}\vspace*{-.5em}
  \caption{Average rate $\bar R$ vs. number of files $N$ ($K=4$, $M=1$ kbit. File size: $F_n=n/N$ kbit, $n\in\Nc$, equal file popularity: $p_n=1/N$, $n\in\Nc$.). }\label{performance_3_RvsN}\vspace*{-1em}
\end{figure}

In  Sections~\ref{sec:simA} and~\ref{sec:simB}, we have focused on how the average rate $\bar{R}$  changes with $M$ and $K$ for  a given set of $N$ files with specific nonuniform popularity distribution.  When all files have uniform popularity but nonuniform file size, we can also evaluate how $\bar{R}$ grows with $N$ under different schemes. In Fig.~\ref{performance_3_RvsN},  we consider equal file popularity  but different file sizes and plot $\bar R$ vs. $N$ for  $K=4$ users and $M=1$ kbit.  We set the size of file $n$ as $n/N$ kbit for $n=1,\ldots,N$. For comparison, we consider a  D-CCS based scheme in~\cite{Zhang&Lin19Closing} and the optimized D-CCS in~\cite{Wang2019Optimization}. For solving {\bf P0} using the successive GP approximation algorithm, we provide the result for $N\le 6$, due to its high computational complexity. Similar to pervious results, the PF-SA-cache-placement-based approach and the successive GP approximation method have nearly identical performance. They achieve the lowest $\bar R$ among all the schemes and are very close to the proposed lower bound. This shows the effectiveness of our proposed algorithms in this case and demonstrates  the optimized D-MCCS has a near-optimal performance. 

Finally, the average computation time of our proposed two algorithms for the D-MCCS in generating Fig.~\ref{performance_3_RvsN} is given in Table~\ref{table3}.  We observe that with as $N$ increases, the average computation time of the successive GP approximation algorithm increases very fast and becomes impractical beyond $N=6$. In contrast, the average  computation time of the PF-SA-cache-placement-based algorithm  remains very low, and the growth rate is very mild as  $N$ increases. 

%%%%%%%%%%%%%%%%%%%%%%%%%%%%%%%%%%%%%%%%%%%%%%%%%%%%%%%
\section{Conclusion and Discussion}\label{sec:conclusion}
In this paper, we studied the memory-rate tradeoff for decentralized caching with nonuniform file popularity and size. Focusing on the D-MCCS, we formulated the cache placement optimization problem and developed two algorithms to solve this non-convex optimization problem: a successive GP approximation algorithm to compute a stationary point and a simple low-complexity PF-SA-cache-placement-based 
scheme, which partitions files into two file groups for cache placement, taking into account the nonuniformity of file popularity and size,  to obtain an approximate solution. We further proposed a lower bound
for decentralized caching. It is given by a non-convex optimization problem, and we adopted the successive GP approximation algorithm to solve it. We showed that the optimized D-MCCS attains the lower bound in some special cases and thus characterizes the exact memory-rate tradeoff. Our numerical study showed that the optimized D-MCCS with our proposed two algorithms in general achieves a near-optimal performance.
Furthermore, our proposed PF-SA-cache-placement-based approximate algorithm maintains a very low computational complexity as $N$ or $K$ increases.

There are several directions for extension  based on this work that can be further explored as future work. First, in this work, our design focused on the average rate performance. The peak rate can be considered for the worst-case scenario to provide additional insights about decentralized caching. Our optimization approach  (\ie {\bf P0} and {\bf P1}) can still be used for the peak rate consideration with some modifications, and a similar algorithm to Algorithm~\ref{alg:MCCS1}  can be constructed to compute a solution. Second, in this work, we focused on the  nonuniformity of files while assuming the cache sizes are the same among users. It will be interesting to extend this work to further consider nonuniform cache sizes. Note that having nonuniform cache sizes  poses new challenges as the cache placement   will now not only be different for files but also depend on each user cache size. This complicates both design and analysis of the coded messages and the evaluation of the average rate. Also, the D-MCCS is only designed for uniform cache size, and the coded delivery scheme needs to be redesigned. 
Finally, we point out that, as discussed in the introduction, besides the delivery schemes used in the D-CCS and the D-MCCS,  several existing works also proposed improved coded delivery schemes    for centralized caching\cite{Ramakrishnan2015An,Wan2017Novel,Asghari:TCOM19,Deng&Dong:COMML23}. It would be interesting to explore these delivery schemes  for decentralized caching and further jointly optimize  the delivery scheme and the decentralized cache placement to reduce the average delivery rate.

\appendices

%%%%%%%%%%%%%%%%%%%%%%%%%%%%%%%%%%%%%%%%%%%%%%%%%%%%%%%%%%%%%%%

\section{Proof of Proposition \ref{pro:upper}}\label{proof_pro:upper}
\IEEEproof
We first derive the delivery rate of the D-MCCS under the PF-SA cache placement in~\eqref{equ:two_groups_rate} for given file request vector $\dbf_\Ac$ and the partition of two file groups $N_1$. We divide all the transmitted coded messages into two different types based on the user subsets they are corresponding to. Recall that $\Ac_1$ is the  set of active users who request files in the first file group $\Nc_1$, and $\Qc^s$ is the set of non-redundant groups of size $s$. Denote $\Qc_1^{s}\triangleq\{\Sc\in\Qc^s,\Sc\cap\Ac_1\ne\emptyset\}$ as the set of those non-redundant groups that contain some users in $\Ac_1$. Also denote $\Qc_2^{s}\triangleq\{\Sc\in\Qc^s,\Sc\cap\Ac_1=\emptyset\}$ as the set of remaining non-redundant groups in $\Qc^s$ that do not include any user in $\Ac_1$; in other words, user subsets in $\Qc_2^{s}$ contain only the users in $\Ac_2$. By definition, we have $\Qc^{s}=\Qc^{s}_1 \cup \Qc^{s}_2$. Accordingly, we can rewrite $R_{\text{MCCS}}(\dbf_\Ac;\qbf)$ in
\eqref{CodedMsgPad} as
\begin{align}\label{CodedMsgPad_0}
R_{\text{MCCS}}&(\dbf_\Ac;\qbf)={\sum_{s=0}^{A-1}\sum_{\Sc\in\Qc_1^{s+1}}|\Cc_\Sc|}+{\sum_{s=0}^{A-1}\sum_{\Sc\in\Qc_2^{s+1}}|\Cc_\Sc|},
\end{align}
where the first term is the total size of   coded messages corresponding to user subsets in $\Qc_1^{s+1}, s=0,\ldots,A-1$ and the second term is the total size of the coded messages corresponding to user subsets in 
$\Qc_2^{s+1},s=0,\ldots,A-1$. 

Now we derive the expressions of the first and second terms in \eqref{CodedMsgPad_0} separately. For user subset $\Sc\in\Qc_1^{s+1}$, based on the PF-SA cache placement in~\eqref{equ:placement}, the size of the corresponding coded message in~\eqref{equ:de_msglength} is given by
\begin{align}%\label{proof_N1_0}
&|\Cc_\Sc|\!=\!\max_{k\in\Sc}\!\bigg(\! \frac{M}{\sum_{n\in\Nc_1}\!F_n}\!\bigg)^{\!s}\!\!\bigg(1-\frac{M}{\sum_{n\in\Nc_1}F_n}\bigg)^{\!\!A-s}\!\!\!\!\!F_{d_k},
\ \Sc\!\in\!\Qc_1^{s+1} \nn
\end{align}
where the maximization is only w.r.t. $F_{d_k}$. Substituting the above into the first term in~\eqref{CodedMsgPad_0} and following the definition of $\Qc_1^{s}$, we have the first term in~\eqref{equ:two_groups_rate}.

For the coded messages corresponding to $\Sc\in\Qc_2^{s+1}$, since all the files requested by the users in $\Ac\backslash\{\Ac_1\}$ are only stored at the server, the size of the corresponding coded message is given by
%\vspace*{-.5em}
\begin{align}\label{proof_N1_1}
|\Cc_\Sc|=
\begin{cases}
F_n,\quad\quad  &s=0,\Sc\in\Qc_2^{s+1},k\in\Sc, n=d_k;\\
0,\quad\quad  &s\ge 1,\Sc\in\Qc_2^{s+1}.
\end{cases}
\end{align}
As a result, the second term in \eqref{CodedMsgPad_0} is given by
\begin{align}\label{proof_N1_2}
\sum_{s=0}^{A-1}\sum_{\Sc\in\Qc_2^{s+1}}|\Cc_\Sc|=\sum_{n\in\dbf_{\Ac_2}}F_n.
\end{align}
Thus, we obtain the expression of the delivery rate of the D-MCCS under the PF-SA cache placement $R^{\text{\tiny PF-SA}}_{\text{MCCS}}(\dbf_\Ac;N_1)$ in \eqref{equ:two_groups_rate}. 

Following the above, the average delivery rate $\bar R^{\text{\tiny PF-SA}}_{\text{MCCS}}(N_1)$ of the D-MCCS under PF-SA cache placement is obtained by taking the expectation of $R^{\text{\tiny PF-SA}}_{\text{MCCS}}(\dbf_\Ac;N_1)$ for all possible active user sets $\Ac\subseteq\Kc$ and all possible file requests $\dbf_\Ac$ for a given active user set $\Ac$, as shown in the expression in \eqref{equ:prop_1}.

%%%%%%%%%%%%%%%%%%%%%%%%%%%%%%%%%%%%%%%%%%%%%%%%%%%%%%%%%%%%%%%%%%%
\section{Proof of Corollary \ref{cor:upper}}\label{proof_cor:upper}
\IEEEproof
To prove Corollary~\ref{cor:upper}, we simplify the delivery rate $R_{\text{MCCS}}(\dbf_\Ac;\qbf)$ in~\eqref{CodedMsgPad_0} using the two-file-group-based placement described in~\eqref{equ:placement_cor}. Based on~\eqref{equ:placement_cor}, for the coded message corresponding to the user subset $\Sc\in\Qc_1^{s+1}$, its size in~\eqref{equ:de_msglength} is given by
\begin{align}\label{proof_N1_0_cor}
\!\!\!\!|\Cc_\Sc|=\left( \frac{M}{N_1F}\right)^{s}\left(1- \frac{M}{N_1F}\right)^{A-s}\!\!F, \ \Sc\in\Qc_1^{s+1}.
\end{align}
The number of user subsets in $\Qc_1^{s+1}$ is the total number of non-redundant groups in $\Qc^{s+1}$ subtracting the number of non-redundant groups that only include users in $\Ac_2$, given by
%\begin{align}\label{proof_N1_2}
$\left( \binom{A}{s+1}-\binom{A-\widetilde N(\dbf_\Ac)}{s+1}\right)-\left(\binom{A_2}{s+1}-\binom{A_0-\widetilde N(\dbf_{\Ac_2})}{s+1}\right)$.
%\end{align}
It can be easily shown that $\binom{A}{s+1}-\binom{A-\widetilde N(\dbf_\Ac)}{s+1}=\sum_{i=1}^{\widetilde N(\dbf_\Ac)}\binom{A-i}{s}$. Thus, the total size of the all the coded messages corresponding to the user subsets in $\Qc_1^{s+1}, s=0,\ldots,A-1$, \ie
the first term in \eqref{CodedMsgPad_0} is given by
%\vspace*{-.5em}
\begin{align}\label{equ:two_groups_rate_0_cor}
\sum_{\Sc\in\Qc_1^{s+1}}|\Cc_\Sc|
&=\left( \sum_{i=1}^{\widetilde N(\dbf_\Ac)}\binom{A-i}{s}-\sum_{i=1}^{\widetilde N(\dbf_{\Ac_2})}\binom{A_2-i}{s}  \right)\nn\\
&\ \cdot\left( \frac{M}{N_1F}\right)^{s}\left(1- \frac{M}{N_1F}\right)^{A-s}\!\!F.
%\cdot\frac{M^s(N_1F-M)^{A-s}}{(N_1)^AF^{A-1}}.
%\end{autobreak}
\end{align}
The size of the coded messages corresponding to user subsets in $\Qc_2^{s+1}, s=1,\ldots,A-1$, \ie the second term in~\eqref{CodedMsgPad_0}, is 
\begin{align}\label{proof_N1_2_cor}
\sum_{s=0}^{A-1}\sum_{\Sc\in\Qc_2^{s+1}}|\Cc_\Sc|=\widetilde N(\dbf_{\Ac_2})F.
\end{align}
Thus,
we obtain the expression of  $R_\text{MCCS}(\dbf_\Ac,N_1)$ in \eqref{equ:pupo_two_groups_rate}.

%%%%%%%%%%%%%%%%%%%%%%%%%%%%%%%%%%%%%%%%%%%%%%%%%%%%%%%%%%%%%%%
\section{Proof of Theorem \ref{thm_bnd}}\label{proof_thm_bnd}

\IEEEproof
The proof follows the genie-based approach used in developing the lower bound for the centralized uncoded cache placement under uniform or nonuniform file popularity~\cite{Yu&Maddah-Ali:TIT2018,Deng&DongMCCS:TIT22} or nonuniform cache sizes~\cite{Ibrahim2019Coded}.
For a given file request vector $\dbf_\Ac$ and the corresponding set of distinct file indices $\Dc_\Ac$, the average delivery rate must satisfy~\cite{Deng&DongMCCS:TIT22}
\begin{align}\label{equ:lemma1Prof_1}
R(\Dc_\Ac;\qbf)\geq\max_{\pi:\Ic_{|\Dc_\Ac|}\rightarrow\Dc_\Ac}\sum_{i=1}^{\widetilde N(\dbf_\Ac)}\sum_{s=1}^{A-1}\binom{A-s}{i}a_{\pi(i),s}
\end{align}
where $a        _{\pi(i),s}$ is the number of bits of file $\pi(i)$ cached exclusively by any user subset $\Sc\in\Ac$ with $|\Sc|=s$. With a decentralized cache placement $\qbf$, from~\eqref{equ:de_subfile}, the number of bits cached by any $s$ active users in $\Ac$ is
$a_{\pi(i),s}=q_{\pi(i)}^{s}(1-q_{\pi(i)})^{A-s}F_{\pi(i)}$.
Substitute this expression into \eqref{equ:lemma1Prof_1}, we have
\vspace*{-.5em}
\begin{align}
R(&\Dc_\Ac;\qbf)\geq\nn\\
&\max_{\pi:\Ic_{|\Dc_\Ac|}\rightarrow\Dc_\Ac}\sum_{i=1}^{\widetilde N(\dbf_\Ac)}\sum_{s=1}^{A-1}\binom{A-s}{i}q_{\pi(i)}^{s}(1-q_{\pi(i)})^{A-s}F_{\pi(i)}\nn
\end{align}
where the right hand side is the lower bound on the delivery rate for a given $\Dc_\Ac$ and $\qbf$ in \eqref{R_lblb}. By averaging $R_\text{lb}(\Dc_\Ac;\qbf)$  over all possible $\Dc_\Ac\subseteq\Nc$ and $\Ac\subseteq\Kc$, we obtain the lower bound on the average rate  $\bar R_\text{lb}(\qbf)$  w.r.t $\qbf$ in \eqref{equ:lb_obj1}. The final lower bound on average rate is obtained by optimizing $\qbf$ to minimize $\bar R_\text{lb}(\qbf)$, which is the optimization problem {\bf P3}. 
\endIEEEproof

%%%%%%%%%%%%%%%%%%%%%%%%%%%%%%%%%%%%%%%%%%%%%%%%%%%%%%%%%%%%%%%
\section{Proof of Theorem~\ref{thm_K2}}\label{proof_thm_K2}
\IEEEproof
To show the equivalence of {\bf P0} and {\bf P3} for $A\le 2$,  it is sufficient to show that $\bar R_\text{MCCS}(\qbf)$ and $\bar R_\text{lb}(\qbf)$ in \eqref{A2_MCCS} and \eqref{A2_lb} are equivalent. Comparing $\bar R_\text{MCCS}(\qbf)$ and $\bar R_\text{lb}(\qbf)$, we only need to examine $R_\text{MCCS}(\dbf_\Ac;\!\qbf)$ and $R_\text{lb}(\Dc_\Ac;\!\qbf)$ in \eqref{CodedMsgPad} and \eqref{R_lblb}.
We consider this for$A=1$ and $A=2$ separately below.

\emph{Case 1:  $A=1$.}
Denote $\Ac=\{u_1\}$. In this case, $R_\text{MCCS}(\dbf_\Ac;\qbf)$ in \eqref{CodedMsgPad} can be straightforwardly rewritten as
\begin{align*}
R_\text{MCCS}(\dbf_\Ac;\qbf)
&=(1-q_{d_{u_1}})F_{d_{u_1}}.
\end{align*}
For $\Ac=\{u_1\}$, we have $\Dc_\Ac=\{d_{u_1}\}$. Thus, $R_\text{lb}(\Dc_\Ac;\qbf)$ in \eqref{R_lblb} is given by
\begin{align}\label{equ_lbMCCS_1_equ_A1}
\!\!R_\text{lb}(\Dc_\Ac;\qbf)\!=\!(1\!-q_{d_{u_1}})F_{d_{u_1}}\!\!=\!\!R_\text{MCCS}(\dbf_\Ac;\qbf),\ |\Ac|=1.
\end{align}

\emph{Case 2: $A=2$.}
Denote $\Ac=\{u_1,u_2\}$. In this case, the two active users can have either the same or distinct file requests. We discuss the two cases below.

\subsubsection{$d_{u_1}=d_{u_2}$} Two users request the same file, and we have $\widetilde{N}(\dbf_{\Ac})=1$.
Without loss of generality, we denote the leader group as $\Uc_\Ac=\{u_1\}$. By Definition~\ref{defRedundant}, the set of non-redundant groups is $\{\{u_1\},\{u_1,u_2\}\}$, and we have  $\Qc^1=\{\{u_1\}\}$ and $\Qc^2=\{\{u_1,u_2\}\}$. Thus, we can rewrite \eqref{CodedMsgPad} as
\begin{align*}
R_\text{MCCS}(\dbf_\Ac;\qbf)&=\sum_{s=0}^{1}\sum_{\Sc\in\Qc^{s+1}}\!\!\max_{k\in\Sc}q_{d_{k}}^{s}(1-q_{d_{k}})^{2-s}F_{d_k}\nn\\
&=(1-q_{d_{u_1}})^2F_{d_{u_1}}+q_{d_{u_1}}(1-q_{d_{u_1}})F_{d_{u_1}}.
\end{align*}
Given the leader group $\Uc_\Ac=\{u_1\}$, we have $\Dc_\Ac=\{d_{u_1}\}$. Thus, $R_\text{lb}(\Dc_\Ac;\!\qbf)$ in \eqref{R_lblb} is
given by\begin{align}\label{equ_lbMCCS_1_equ}
R_\text{lb}(\Dc_\Ac;\!\qbf)&=(1-\!q_{d_{u_1}}\!)^2\!F_{d_{u_1}}+q_{d_{u_1}}\!(1-q_{d_{u_1}}\!)F_{d_{u_1}}\nn\\
&=R_\text{MCCS}(\dbf_\Ac;\qbf).
\end{align}

\subsubsection{$d_{u_1}\ne d_{u_2}$} When two active users request different files, we have $\widetilde{N}(\dbf_{\Ac})=2$. The leader group is  $\Uc_\Ac=\{u_1,u_2\}$, and the set of non-redundant groups is $\{\{u_1\},\{u_2\}\{u_1,u_2\}\}$, of which can be categorized as  $\Qc^1=\{\{u_1\},\{u_2\}\}$ and $\Qc^2=\{\{u_1,u_2\}\}$. Thus, $R_\text{MCCS}(\dbf_\Ac;\qbf)$ in \eqref{CodedMsgPad} is given by
\begin{align}
R_\text{MCCS}(\dbf_\Ac;\qbf)\nn=&\sum_{s=0}^{1}\sum_{\Sc\in\Qc^{s+1}}\!\!\max_{k\in\Sc}q_{d_{k}}^{s}(1-q_{d_{k}})^{2-s}F_{d_{k}}\nn\\
=&(1-q_{d_{u_1}})^2F_{d_{u_1}}+(1-q_{d_{u_2}})^2F_{d_{u_2}}\nn\\
+\max&\{q_{d_{u_1}}(1-q_{d_{u_1}})F_{d_{u_1}},q_{d_{u_2}}(1-q_{d_{u_2}})F_{d_{u_2}}\}.\nn
\end{align}
Also, $R_\text{lb}(\Dc_\Ac;\qbf)$ in \eqref{R_lblb} is
given by\begin{align}\label{equ_lbMCCS_2_equ}
&R_\text{lb}(\Dc_\Ac;\qbf)= \nn\\
&\max\big\{\!(1\!\!-q_{d_{u_1}}\!)^2F_{d_{u_1}}\!\!+\!(1\!-\!q_{d_{u_2}})^2F_{d_{u_2}}\!+q_{d_{u_1}}(1\!-q_{d_{u_1}})F_{d_{u_1}}\!,\nn\\
%&(1-q_{d_{u_1}})^2+(1-q_{d_{u_2}})^2+q_{d_{u_2}}(1-q_{d_{u_2}})\big\}\nn\\
&\quad \quad\ (1\!\!-\!q_{d_{u_1}})^2F_{d_{u_1}}\!\!+(1\!-\!q_{d_{u_2}})^2F_{d_{u2}}\!+q_{d_{u_2}}(1\!-q_{d_{u_2}})F_{d_{u_2}}\!\}\nn\\
&\quad\quad \quad\quad\ \ =R_\text{MCCS}(\dbf_\Ac;\qbf).
\end{align}
From \eqref{equ_lbMCCS_1_equ} and \eqref{equ_lbMCCS_2_equ}, we conclude  $\bar R_\text{MCCS}(\qbf)=\bar R_\text{lb}(\qbf)$ for $A=2$.
Combining Cases 1 and 2, we can conclude that $\bar R_\text{MCCS}(\qbf)=\bar R_\text{lb}(\qbf)$ for $A\le2$.
\endIEEEproof

\section{Proof of Proposition~\ref{prop_1}}\label{proof_prop_2}
\IEEEproof
For $F=F_1=\ldots=F_N$ and $q^*_1=\cdots=q^*_N$, the size of coded message $\Cc_\Sc$ corresponding to $\Sc\in\Qc^{s+1}$  in~\eqref{equ:de_msglength} is given by
\begin{align}
|\Cc_\Sc|=\max_{k\in\Sc}(q_{d_{k}}^*)^{s}(1-q_{d_{k}}^*)^{A-s}F=(q_{1}^{*})^{s}(1-q_{1}^{*})^{A-s}F.\nn
\end{align}
Following this, $R_{\text{MCCS}}(\dbf_\Ac;\qbf^*)$ in \eqref{CodedMsgPad} is given by
\begin{align}\label{equ:propo_1}
\!\!\!R_{\text{MCCS}}(\dbf_\Ac;\qbf^*)&=\sum_{s=0}^{A-1}\sum_{\Sc\in\Qc^{s+1}}(q_{1}^{*})^{s}(1-q_{1}^{*})^{A-s}F
\end{align}
where the summation is over all the non-redundant groups. By Definition~\ref{defRedundant}, the number of non-redundant groups in $\Qc^{s+1}$ is $\binom{A}{s+1}-\binom{A-\widetilde N(\dbf_\Ac)}{s+1}=\sum_{i=1}^{\widetilde{N}(\dbf_\Ac)}\binom{A-i}{s}$. Thus, we can rewrite \eqref{equ:propo_1} as
\begin{align}
R_{\text{MCCS}}(\dbf_\Ac;\qbf^*)&=\sum_{s=0}^{A-1}\sum_{i=1}^{\widetilde{N}(\dbf_\Ac)}\binom{A-i}{s}(q_{1}^{*})^{s}(1-q_{1}^{*})^{A-s}F\nn\\
&=R_{\text{lb}}(\Dc_\Ac;\qbf^*).
\end{align} Thus, we conclude that
$\bar R_{\text{MCCS}}(\qbf^*)=\bar R_{\text{lb}}(\qbf^*)$.
\endIEEEproof

%\balance
\bibliographystyle{IEEEtran}
\bibliography{Yong,IEEEabrv}

\end{document}